\documentclass[twocolumn]{aa} 

\usepackage{graphicx}
\begin{document}
   \title{Chemical enrichment and star formation in the Milky Way disk}
   \subtitle{III. Chemodynamical constraints}

   \author{H. J. Rocha-Pinto
           \inst{1,4} 
           Chris Flynn
           \inst{2}
           John Scalo
           \inst{3}
           J. H\"anninen
           \inst{2}
           W. J. Maciel
           \inst{1}
           \and 
           Gerhard Hensler
           \inst{5} }

   \offprints{W. J. Maciel}

   \institute{Depart. of Astronomy, IAG/USP, 
              Rua do Mat\~ao 1226, 05508-900 S\~ao Paulo SP, 
              Brazil. \email{maciel@astro.iag.usp.br} 
             \and Tuorla Observatory, V\"ais\"al\"antie 20, FI-21500, Pikkii\"o, Finland. 
              \email{cflynn@astro.utu.fi, jyrki@astro.utu.fi}
             \and Depart. of Astronomy, The University of Texas at Austin, 
              USA. \email{parrot@astro.as.utexas.edu}
            \and
              Depart. of Astronomy, University of Virginia, P.O. Box 3818, Charlottesville, 
              VA 22903-0818, USA. \email{helio@virginia.edu}
            \and
              Institut f\"ur Theoretische Physik und Astrophysik, Universit\"at 
              Kiel, D-24098 Kiel, Germany. \email{hensler@astrophysik.uni-kiel.de}  
             }

   \date{Received ; accepted }

   \abstract{In this paper, we investigate some chemokinematical properties of the Milky Way 
disk, by using a sample composed by 424 late-type dwarfs. We show that the velocity dispersion 
of a stellar group correlates with the age of this group, according to a law proportional 
to  $t^{0.26}$, where $t$ is the age of the stellar group. The temporal evolution of the vertex 
deviation is considered in detail. It is shown that the vertex deviation does not seem to depend 
strongly on the age of the stellar group. Previous studies in the literature seem to not have found 
it due to the use of statistical ages for stellar groups, rather than individual ages. The 
possibility to use the orbital parameters of a star to derive information about its birthplace is 
investigated, and we show that the mean galactocentric radius is likely to be the most reliable 
stellar birthplace indicator. However, this information cannot be presently used to derive radial 
evolutionary constraints, due to an intrinsic bias present in all samples constructed from nearby 
stars. An extensive discussion of the secular and stochastic heating mechanisms commonly invoked to 
explain the age--velocity dispersion relation is presented.  We suggest that the age--velocity 
dispersion relation could reflect the gradual decrease in the turbulent velocity dispersion from 
which disk stars form, a suggestion originally made by Tinsley \& Larson (\cite{tinsley}) and 
supported by several more recent disk evolution calculations.  A test to distinguish between 
the two types of models using high-redshift galaxies is proposed.
   }

   \authorrunning {H.J. Rocha-Pinto et al.}
   \titlerunning {Chemical enrichment and star formation III.}

   \maketitle
%

\section{Introduction}

The publication of the data from HIPPARCOS (ESA \cite{ESA}) improved substantially kinematical 
studies of the solar neighbourhood. Simultaneous measurements of parallaxes and proper motions, 
made by the astrometric satellite, allow the calculation of tangential velocities for tens of 
thousands of stars. In spite of it, a complete study of the stellar velocity distribution in 
the vicinity of the Sun is yet to be made, since the database for radial velocities has not 
increased in the same proportion, and is still a major problem in stellar kinematic studies. 

The correlation between kinematical and chemical properties of the disk stars is a classical 
result of galactic astronomy (Roman \cite{roman50}, \cite{roman52}; Eggen, Lynden-Bell \& Sandage 
\cite{eggen}; Edvardsson et al. \cite{Edv93}, hereafter Edv93). The knowledge about both the 
kinematical and chemical properties of the stellar populations gives several clues about the 
evolution of a galaxy, unveiling the complex processes that concurred in its building. 

Recently, we have built a large sample having chromospheric ages and photometric metallicities. 
This sample was used to study the age--metallicity relation (Rocha-Pinto et al. \cite{paper1}, 
hereafer Paper I) and the star formation history (Rocha-Pinto et al. \cite{paper2}, hereafter 
Paper II). It is interesting to use these data in the study of chemodynamical constraints. Our 
sample is small, compared to that used in some recent investigations, but it can still be useful 
for the investigation of some important constraints. Several papers that address constraints between 
stellar ages and kinematical properties make use of ages calculated by statistical methods, that is, 
average ages for a group of stars. The advantage of our approach is that we consider absolute ages, 
and the constraints are expected to give a more accurate description of the galactic kinematics. 

This paper is organized as follows. The kinematical sample is described in detail in Section~2. The 
age--velocity dispersion relation (hereafter, AVR) is derived in Section~3, where we show that all 
components of the spatial velocity of the star present a similar AVR. In Section~4, we investigate 
the temporal evolution of the vertex deviation. Section~5 investigates the possibility of using 
orbital parameters to find the stellar birthplace and age. In Section~6, we attempt to use orbital 
information to derive some radial constraints to chemical evolution parameters. The AVR is readdressed 
in Section~7, where we present a thorough discussion of its interpretation, concluding that secular 
disk heating mechanisms are incapable of accounting for the large dispersions of old disk stars, 
and suggesting that a major component of the AVR may be due to disk cooling, as originally 
suggested by Tinsley \& Larson (1978). Our conclusions are presented in Section 8.

\section{The kinematic sample}

We have looked for kinematical data in the literature, for the 729 stars of the initial sample 
and for the 261 of the aditional sample (see Paper I for details). Proper motions were obtained 
from the HIPPARCOS catalogue or the SIMBAD database. Radial velocities come from several sources:
Barbier-Brossat, Petit \& Figon (1994), Duflot et al. (\cite{duflota}, \cite{duflotb}), 
Fehrenbach et al. (\cite{fehren96}, \cite{fehren97}), 
Nordstr\"om et al. (\cite{NSLA}) and in the SIMBAD database. In total, 459 stars
have radial velocities in the literature, from which 424 have chromospheric
ages lower than 15 Gyr. In the following, we will use only these 424 stars.

We calculated the spatial velocities $U$, $V$ and $W$, in a system pointing to the galactic 
anticenter, the direction of the galactic rotation and the direction towards the north galactic 
pole, respectively, using the equations of Johnson \& Soderblom (1987). The velocities are further 
transformed into peculiar velocities with respect to the local standard of rest, $u$, $v$ and $w$, 
using the {\sl basic solar motion}\footnote{The value of $v_\odot$ is still uncertain. Dehnen \& 
Binney (\cite{dehbin}) have calculated $v_\odot = 5.25$ km/s, but Feast (\cite{feast}) argues 
that several recent determinations give $v_\odot \sim 11$ km/s. We decided to use this more 
conservative value in our analysis.} (Mihalas \& Binney 1981), 
$(u_\odot, v_\odot, w_\odot) = (-9, 11, 6)\, {\rm km}\,{\rm s}^{-1}$.

    \begin{table*} 
      \caption[]{Solar neighbourhood late-type dwarfs (complete table is only available through 
       electronic ftp)} 
         \label{datall} 
         \begin{flushleft} 
    {\halign{%
    #\hfil &  \quad \hfil$#$\hfil & \quad \hfil$#$  &  
    \quad \hfil$#$  & \quad\hfil $#$\hfil &  
     \quad\hfil $#$   
    & \quad\hfil $#$ &  \quad\hfil $#$ & \quad\hfil $#$\hfil & \quad\hfil $#$\hfil & \quad\hfil $#$\hfil 
    & \quad\hfil $#$\hfil &  \quad\hfil $#$\hfil 
    & \quad\hfil $#$\hfil &  \quad\hfil $#$\hfil \cr  
    \noalign{\hrule\medskip} 
    HD/BD  & \log R'_{HK} & {\rm [Fe/H]} & v_r & {\rm Age} & u & v & w & \varepsilon_u & \varepsilon_v 
    & \varepsilon_w
    & R_m & R_p & Z_{\rm max} & e \cr  
        &    & {\rm dex} & {\rm km/s} & {\rm Gyr} & {\rm km/s} & {\rm km/s} & {\rm km/s} & {\rm km/s} 
    & {\rm km/s} & {\rm km/s} &
    {\rm kpc} & {\rm kpc} & {\rm kpc} & \cr
    \noalign{\medskip\hrule\medskip} 
166  &  -4.33 &  0.22 &  -7.5 &  0.17 &   6 & -11 &  -3 & 2 & 4 & 3 &      &      &      &      \cr				
400  &  -4.95 & -0.35 & -13.8 &  8.54 & -36 &   2 &  -3 & 2 & 4 & 2 & 8.54 & 7.47 & 0.02 & 0.13 \cr
693  &  -4.96 & -0.56 &  14.4 & 14.32 & -28 &  -2 & -12 & 1 & 1 & 5 & 8.36 & 7.57 & 0.15 & 0.09 \cr
1320 &  -4.87 & -0.30 &   9.3 &  5.69 &  47 & -30 & -10 & 2 & 1 & 5 & 6.41 & 0.71 & 6.04 & 0.89 \cr
1581 &  -4.84 & -0.26 &   9.1 &  4.58 &  64 &   7 & -38 & 2 & 2 & 4 & 7.54 & 6.34 & 0.12 & 0.16 \cr
1835 &  -4.44 &  0.28 &  -2.7 &  0.21 &  27 &  -3 &   6 & 1 & 1 & 5 & 8.95 & 7.10 & 0.75 & 0.21 \cr
2151 &  -5.00 &  0.00 &  22.7 &  3.57 &  52 & -36 & -24 & 2 & 3 & 3 & 8.31 & 7.61 & 0.13 & 0.08 \cr
2913 &  -4.08 & -0.05 &  16.5 &  0.18 &  11 &  11 &  -8 & 2 & 3 & 4 & 7.42 & 6.05 & 0.37 & 0.18 \cr
3229 &  -4.58 & -0.36 &   6.2 &  2.43 &  17 & -18 &  -9 & 1 & 2 & 4 & 7.80 & 7.38 & 0.11 & 0.05 \cr
3651 &  -4.85 &  0.01 & -34.2 &  2.07 & -50 &  -9 &  15 & 2 & 3 & 3 & 8.24 & 6.97 & 0.28 & 0.15 \cr
3765 &  -4.91 & -0.20 & -63.0 &  4.87 & -28 & -71 & -21 & 2 & 4 & 2 & 6.45 & 4.75 & 0.30 & 0.26 \cr
  \noalign{\medskip\hrule}}} 
         \end{flushleft} 
   \end{table*}

The sample is given in Table \ref{datall}, available at the CDS\footnote{
{\tt http://cdsweb.u-strasbg.fr/}}, 
where we list, respectively, the HD number (or BD number for a few stars), 
the chromospheric index $\log R'_{\rm HK}$, the photometric metallicity [Fe/H], 
the radial velocity, the chromospheric age, calculated according to the formalism 
described in Paper I, the peculiar velocities $u$, $v$ and $w$ with their measurement
errors, and some parameters referred to later on.

      \begin{figure*}
      \resizebox{\hsize}{!}{\includegraphics{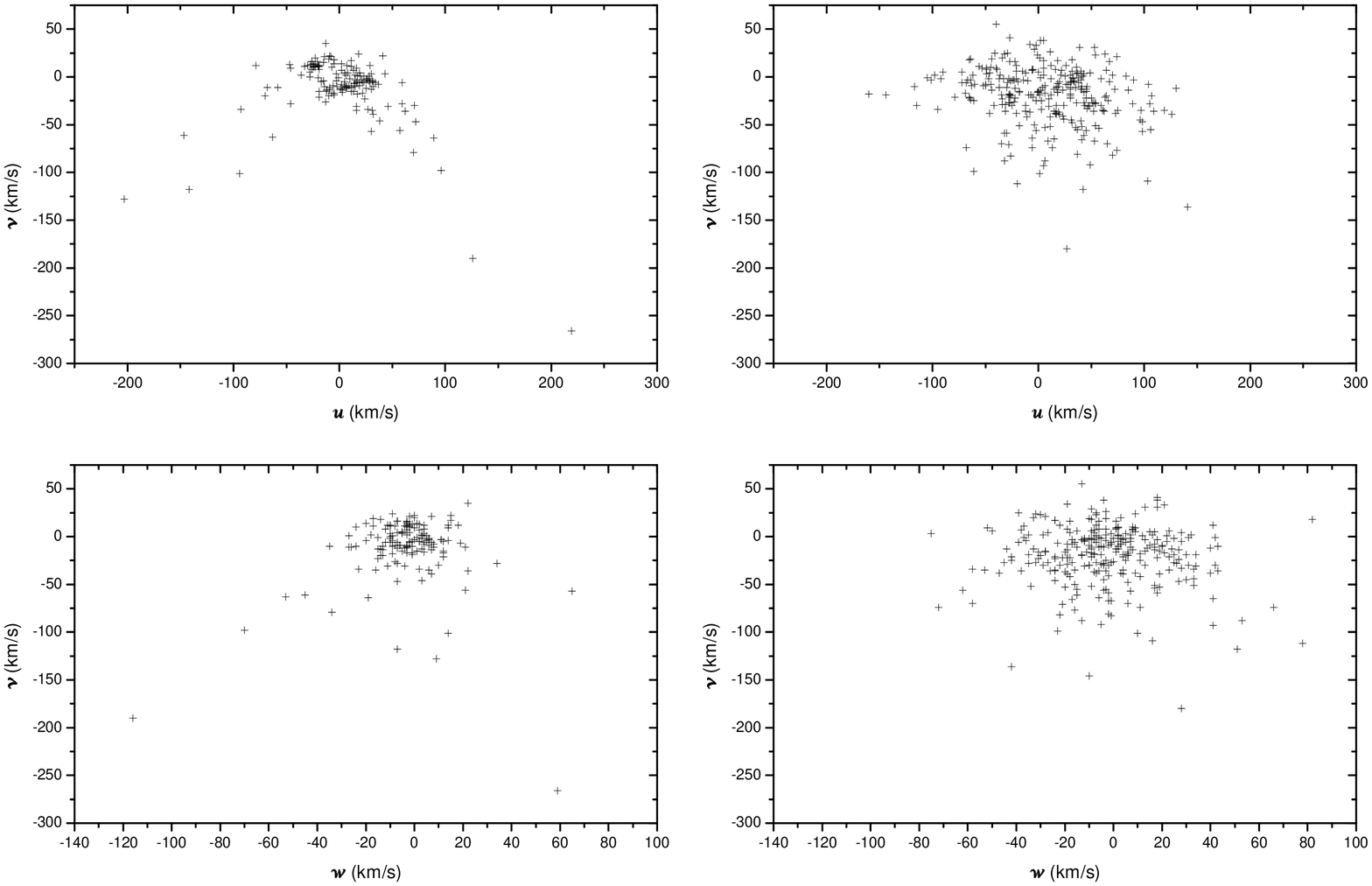} }
      \caption[]{
         Velocity diagrams $v\times u$ and $v\times w$ for the active stars 
        (left panels) and inactive stars (right panels). Compare the increase 
        in the velocity dispersion of the inactive stars with that of the active stars. 
        Note that, in spite of their low velocity dispersion, amongst active stars there are 
        objects with velocity components typical of an old population (Rocha-Pinto et al. \cite{cyko}).  
          }
      \label{act}
      \end{figure*}

Spatial velocity distributions are strongly dependent on the mean age of the stellar group studied. 
We can consider stellar subgroups having different ages, to investigate their velocity distribution.  
The most direct division in our sample is set by the chromospheric activity levels, since it 
was constructed from chromospheric activity surveys. Active and inactive stars must present different 
velocity distributions, that will reflect the mean ages of these subgroups. The active stars are 
expected to be more concentrated in the velocity space, compared to the inactive stars, as shown 
by Soderblom (\cite{soder90}) for a sample of 32 active stars. The same result was found by Jeffries 
\& Jewell (1993), from a sample of dwarfs detected in the extreme ultraviolet by ROSAT.

In Fig. \ref{act}, we show the velocity diagrams for these two subgroups of chromospheric activity. 
The diagrams  for the active stars are presented in the left panels, and those for the inactive stars 
in the right panels. The plot  confirms the previous conclusions. Active stars present a low velocity 
dispersion, that is explained by their youth, compared to the inactive stars. Note that 
some active stars present velocity components typical of an advanced age. Soderblom (\cite{soder90}) 
has already pointed to this problem. However, in his sample of 32 dwarfs, only 3 showed high velocity 
components. Rocha-Pinto \& Maciel (\cite{RPM98}, hereafter RPM98) have increased this number to 10. The 
present sample contains 29 stars like these. Jahrei\ss, Fuchs \& Wielen (1999) also mention the existence of 
chromospherically active high velocity stars. Comparing isochrone, chromospheric and lithium ages 
of these objects, Rocha-Pinto et al. (\cite{cyko}) concluded that they can be single stars formed out 
of the coalescence of short-period binaries. The coalescence makes the star look chromospherically younger, 
because the orbital angular momentum of the former binary is transformed into rotational angular momentum 
of the newly formed single star; however, since the binary formed some Gyr before the coalescence, 
the movement of its center of mass around the Galactic center, later inherited by the single star, shows 
the kinematical signatures of an advanced age. We call these stars CYKOS (chromospherically young, 
kinematically old stars). 

\section{Age--velocity dispersion relation}

The velocity dispersion of a stellar group is larger, the older is the group. This relation is 
usually interpreted as a result of gravitational perturbations experienced by the stars
during their translations around the galactic center. In spite of the fact that several processes 
which can cause an increase in the velocity dispersion are relatively well known, it is not 
completely understood which of them is the main responsible mechanism in this heating, and this 
comes mainly from the disagreement between the observed dispersions amongst disk dwarfs found by 
different studies.  Several authors have studied this issue (Wielen \cite{wie74}; 
Mayor \cite{may74}; Cayrel de Strobel \cite{cayrel}; Carlberg et al. \cite{carlberg}; Knude et al. 
\cite{knude}; Str\"omgren \cite{stromgren}; Meusinger et al. \cite{meusinger}; Wielen et al. 
\cite{wie92}; Dehnen \& Binney \cite{dehbin}; Fuchs et al. \cite{fuchsetal}) but their results 
are quite conflicting. It is not clear whether this was caused by selection effects, a problem 
which plagues severely kinematical studies, or is an intrinsic property of different 
stellar classes used by them (see Fridman, Khoruzhii \& Piskunov \cite{fridman}).

      \begin{figure}
      \resizebox{\hsize}{!}{\includegraphics{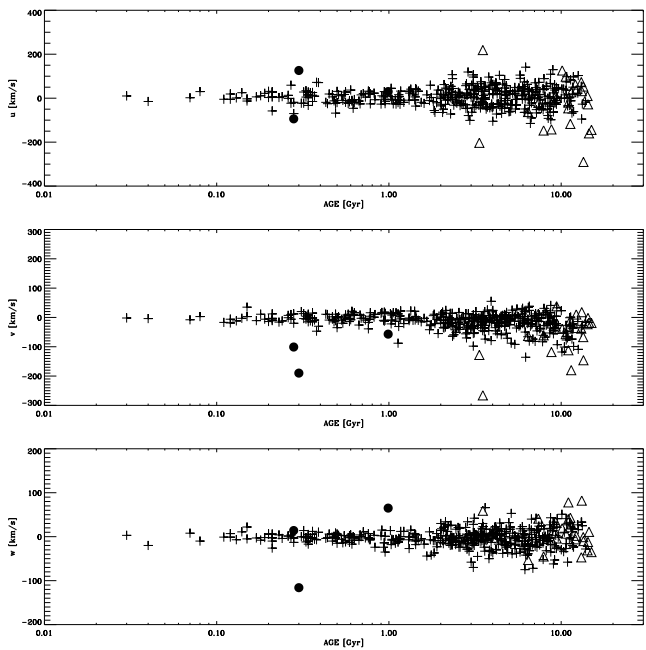} }
      \caption[]{
        Spatial velocities as a function of age. The disk heating can be clearly seen in this 
        picture. Some stars, marked in the plot by solid circles and empty triangles, were 
        discarded from the sample, before the final calculations of the age--velocity dispersion 
        relation (see text).
          }
      \label{vecgrow}
      \end{figure}

In Fig. \ref{vecgrow}, we present plots of peculiar velocities as a function of the stellar age. 
It can be seen clearly that the number of high-velocity stars grows towards older ages. In the 
middle panel, $v$ clearly shows the asymmetric drift (Binney \& Merrifield \cite{merrifield}), 
that is, the older stars rotate around the galactic center with slower velocities with 
respect to the local standard of rest. 

In order to calculate the velocity dispersion we had to further examine our stellar sample. 
First of all, there are three clear outliers, as seen in Fig. \ref{vecgrow} (solid circles). 
For these outliers at least one of the velocity components is not typical of thin disk
stars. These are all CYKOS, which are not supposed to be real young stars, thus, we have 
discarded them from this analysis. Furthermore, we have also excluded from the
velocity dispersion calculations all the stars with metallicity [Fe/H] $ < -0.5$, since they are 
likely to belong to the thick disk.  This condition excludes 24 stars (empty triangles in Fig.2), 
so we are left with 397 stars (crosses) from the 424 star sample. 

Each component of the velocity dispersion is calculated in equal star number bins instead of equal
age range bins. Because the sample stars are not evenly distributed in age we have used bins 
with equal number of stars ($N_{\rm stars/bin} = 40$). This way we have an equal weight for 
each data point when fitting the AVR.

The AVRs that we have found can be approximated by power laws of the kind 
$\sigma\propto (1 +t/\tau)^x$, where $\tau$ is a time scale for the growth of the velocity 
dispersion and $t$ is the stellar age. The exponent $x$ is $0.26\pm 0.01$, $0.24\pm 0.02$, 
$0.30\pm 0.03$ and $0.34\pm 0.02$, for $\sigma_{\rm tot}$, $\sigma_u$, $\sigma_v$ and 
$\sigma_w$, respectively. These exponents can be  considered equal within $2\sigma$. 
The relation in $w$ is slightly steeper.

Nevertheless, this fitting is very sensitive to the last age bin. If it is removed, we find 
$x\approx 0.31$ for $\sigma_{\rm tot}$. If, on the other hand, we use bins equally spaced in age, 
we find $x$ equal to $042\pm 0.04$, $0.40\pm 0.02$, $0.48\pm 0.04$ and $0.41\pm 0.03$, for 
$\sigma_{\rm tot}$, $\sigma_u$, $\sigma_v$ and $\sigma_w$, respectively. The large difference between 
these exponents and those calculated for equal star number bins reinforces our concerns about 
the sensitivy of the observed AVR. The equal star number binning was chosen 
to avoid many of the uncertainties that arise from the use of age range bins with poor statistics.

Other fitting laws commonly used in the literature were tried. 
For a parametrization like $\sigma = a_0 + a_1 t^x$, we have 
found  $x_{\rm tot} = 0.24  \pm 0.04$, $x_u   = 0.23  \pm 0.04$, $x_v   = 0.37  \pm 0.07$ and 
$x_w   = 0.30  \pm 0.09$. For $\sigma \propto t^x$, we have 
$x_{\rm tot} = 0.26  \pm 0.03$, $x_u   = 0.23  \pm 0.04$, $x_v   = 0.28  \pm 0.04$ and 
$x_w   = 0.35  \pm 0.06$. 

\begin{figure*} 
      \centering
      \includegraphics[width=17 cm]{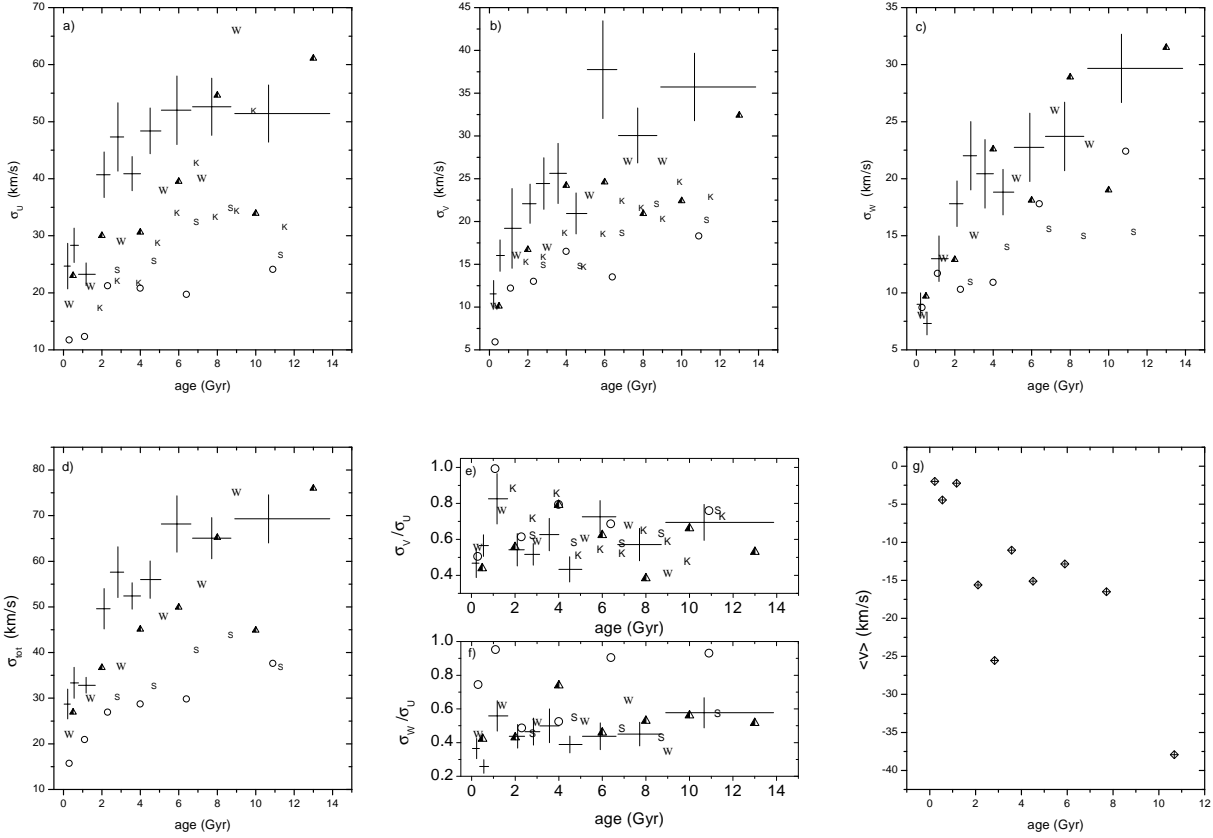}  
       \caption[]{ 
        Characterization of the kinematical sample with a velocity ellipsoid. The panels a,  
        b, c and d stand for the relations between the age and the velocity dispersions in the  
        components $u$, $v$, $w$, and for the total space velocity ($\sigma^2_{\rm tot}=\sigma^2_u+
        \sigma^2_v+\sigma^2_w$), respectively. The velocity dispersions were calculated with respect 
        to the principal axis of the  velocity ellipsoid for each age group. Panels e and f show 
        the evolution of the axial ratios $\sigma_v/\sigma_u$ and $\sigma_w/\sigma_u$ of the 
        ellipsoid. Panel g shows the asymmetrical drift as a function of age.  In these plots the 
        symbols indicate: error bars, our data, calculated by binning the stars in intervals of equal 
        star number ($N = 40$ stars), without applying the $|W|$ weights; 
        semifilled triangles, Cayrel de Strobel \cite{cayrel};  
        letters W, Wielen 1974; letters K, Knude et al. \cite{knude}; letters S, Str\"omgren 1987; 
        empty circles, Meusinger et al. 1991. 
          } 
      \label{agevelc} 
      \end{figure*}

Our age--velocity dispersion relations are given in Fig. \ref{agevelc}, which shows that
our dispersions are considerably higher than those found by Knude et al. (\cite{knude}), 
Str\"omgren (1987) and Meusinger et al. (\cite{meusinger}), for the older ages. 
The agreement is substantially better for the earlier studies by Wielen (\cite{wie74}) and Cayrel 
de Strobel (\cite{cayrel}). The data given by Fuchs et al. (\cite{fuchsetal}) follow closely those 
of Wielen (\cite{wie74}), and are not shown in the figure.

The data in Fig. \ref{agevelc} are presented also in Table \ref{vecdisp}. In this 
table, the first column gives the age range considered, and the other columns give, respectively, 
$\sigma_{\rm tot}$ and the error in this quantity; $\sigma_u$, $\sigma_v$, $\sigma_w$, 
and their respective errors, each shown after the corresponding velocity dispersion; the axial 
ratios $\sigma_v:\sigma_u$ and $\sigma_w:\sigma_u$; and the asymmetric drift 
according to the age of the stellar group.

Although our data agrees at first glance with those given by Fuchs et al. (\cite{fuchsetal}), 
the fitted velocity dispersion laws in their work and ours differ considerably. It is important 
to understand the cause of this difference. Fuchs et al. have used a sample of late-type 
stars based on chromospheric ages, like ours. But they have not taken in consideration 
the metallicity dependence of the chromospheric indices (Rocha-Pinto \& Maciel {\cite{RPM98}), 
when calculating the stellar ages. Moreover, they have weighted each star by the modulus of the 
vertical velocity, $\vert W\vert$, before calculating the velocity dispersions. 
The $\vert W\vert$ weighting was introduced by Wielen (\cite{wie74}) to take into account 
the probability for the detection of each star in the local sample, given their vertical 
oscillation with respect to the Galactic plane. 

By disregarding the metallicity dependence of the chromospheric ages, Fuchs et al. (\cite{fuchsetal}) 
are likely to mix hotter older stars with somewhat `warm' intermediate-age stars. This could boost 
the velocity dispersion to high values at 4-6 Gyr, resulting in a steeper AVR. 
This is illustrated more properly in Fig. \ref{withwithout}, for $u$. We have 
investigated this effect with our sample. The chromospheric age was recalculated 
using the calibration given by Donahue (\cite{don}), that is, without any correction 
for the metallicity of the star. The stars were binned by groups of 40, as in the former cases. 
For $\sigma \propto t^x$, we have 
$x_{\rm tot} = 0.33  \pm 0.05$, $x_u   = 0.29  \pm 0.05$, $x_v   = 0.39  \pm 0.05$ and 
$x_w   = 0.37  \pm 0.08$. This result shows that by disregarding the [Fe/H] dependence in the chromospheric
ages, the exponents of the AVR have increased by 6\% to 39\% from the supposedly real values. 
This is an important result, since it can shed some light into the different slopes found by 
different groups, using different star groups. As Fridman et al. (\cite{fridman}) point out, there are 
two main points of view in the literature for the AVR exponent: Wielen and colleagues 
have found higher values than Knude et al. (\cite{knude}), 
Str\"omgren (1987) and Meusinger et al. (\cite{meusinger}). The results by the Heidelberg group 
are largely based on chromospheric age estimates that disregard the metallicity dependence of 
the chromospheric ages, while the later studies are based on isochrone ages which are free of 
this systematic error. Since our results use a corrected cromospheric age scale, we should expect 
to find exponents closer to those found from isochrone age studies.

    \begin{table*} 
      \caption[]{Velocity dispersions, axial ratios and asymmetrical drift.} 
         \label{vecdisp} 
         \begin{flushleft} 
    {\halign{%
    \hfil #\hfil &  \qquad\hfil $#$\hfil & \qquad\hfil $#$\hfil  &  
    \qquad\hfil $#$\hfil  & \qquad\hfil $#$\hfil &  
     \qquad\hfil $#$\hfil   
    & \qquad\hfil $#$\hfil &  \qquad\hfil $#$  \cr  
    \noalign{\hrule\medskip} 
    $\langle {\rm Age (Gyr)} \rangle$ & \sigma_{\rm tot}  
    & \sigma_u &  
    \sigma_v & \sigma_w & \sigma_v:\sigma_u & 
    \sigma_w:\sigma_u & \langle V \rangle \cr
    \noalign{\medskip\hrule\medskip} 
 0.22 &  29\pm 3 & 25\pm 4 & 12\pm 2 &  9\pm 1 & 0.47\pm 0.08 & 0.36\pm 0.06 & -2 \cr 
 0.55 &  33\pm 3 & 28\pm 3 & 16\pm 2 &  7\pm 1 & 0.57\pm 0.06 & 0.26\pm 0.04 & -4 \cr 
 1.17 &  33\pm 3 & 23\pm 2 & 19\pm 5 & 13\pm 2 & 0.83\pm 0.14 & 0.55\pm 0.09 & -2 \cr 
 2.11 &  50\pm 4 & 41\pm 4 & 22\pm 2 & 18\pm 2 & 0.54\pm 0.09 & 0.43\pm 0.07 & -16 \cr 
 2.84 &  57\pm 6 & 47\pm 6 & 24\pm 3 & 22\pm 3 & 0.52\pm 0.06 & 0.46\pm 0.08 & -26 \cr 
 3.57 &  52\pm 3 & 41\pm 3 & 25\pm 4 & 20\pm 3 & 0.63\pm 0.09 & 0.49\pm 0.10 & -11 \cr 
 4.51 &  56\pm 3 & 48\pm 4 & 21\pm 2 & 18\pm 2 & 0.43\pm 0.07 & 0.39\pm 0.05 & -15 \cr 
 5.90 &  68\pm 4 & 52\pm 6 & 38\pm 6 & 23\pm 3 & 0.73\pm 0.09 & 0.44\pm 0.08 & -13 \cr
 7.71 &  65\pm 4 & 53\pm 5 & 30\pm 3 & 24\pm 3 & 0.57\pm 0.09 & 0.45\pm 0.07 & -17  \cr
10.67 &  69\pm 4 & 51\pm 5 & 36\pm 4 & 30\pm 3 & 0.69\pm 0.10 & 0.58\pm 0.09  & -38 \cr
  \noalign{\medskip\hrule}}} 
         \end{flushleft} 
   \end{table*}

      \begin{figure}
      \resizebox{\hsize}{!}{\includegraphics{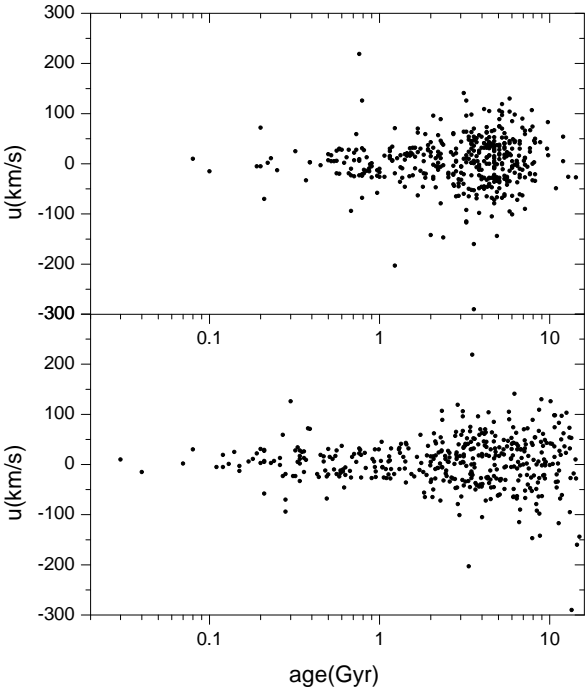} }
      \caption[]{ 
        Increase of the $u$ peculiar velocity with age, for uncorrected (panel a) 
        and corrected (panel b) chromospheric ages. 
        The effect of neglecting the [Fe/H] dependence of the chromospheric age 
        calibration is seen in {\it panel a}, where the mixing of old hotter stars 
        with intermediate-age stars boosts the velocity dispersion around 4-6 Gyr, 
        yielding a steeper AVR, which otherwise would present a slower 
        increase (panel b).
          }
      \label{withwithout}
      \end{figure}

The $\vert W\vert$ weighting also affects considerably the resulting AVR. 
For $\sigma \propto t^x$, and using the uncorrected chromospheric ages, as above, as in 
Fuchs et al. (\cite{fuchsetal}), we have found 
$x_{\rm tot} = 0.50  \pm 0.08$, $x_u   = 0.36  \pm 0.09$, $x_v   = 0.49  \pm 0.13$ and 
$x_w   = 0.53  \pm 0.07$. Qualitatively similar values are found when we use the $\vert W\vert$ weighting 
with the corrected ages and our preferred heating law $\sigma\propto (1 +t/\tau)^x$: 
$x_{\rm tot} = 0.50  \pm 0.02$, $x_u = 0.47  \pm 0.02$, $x_v = 0.57  \pm 0.04$ and 
$x_w   = 0.46  \pm 0.03$. However, these fits are somewhat worse than the previous ones for which no 
vertical weight was applied. 

    \begin{table} 
      \caption[]{Properties of the best-partition groups.} 
         \label{partition} 
         \begin{flushleft} 
    {\halign{%
    #\quad & \hfill $#$\hfil &  \quad\hfil $#$\hfil & \quad\hfil $#$\hfil  &  \quad\hfil $#$\hfil  
    & \quad\hfil $#$\hfil &  
     \quad\hfil $#$\hfil  & \quad\hfil $#$\hfil  & \quad\hfil $#$\hfil & \quad\hfil #\hfil \cr  
    \noalign{\hrule\medskip} 
      group & \overline{\rm [Fe/H]} & \overline{u} & \overline{v}  &  \overline{w}  & \sigma_u &  
     \sigma_v  & \sigma_w & \overline{\rm Age}& number \cr  
    \noalign{\medskip\hrule\medskip} 
    I  & -0.07 & 5 &  -9 & -3 & 38 & 20 & 16 & 3.5 & 359 \cr
    II & -0.37 & 4 & -55 & -3 & 84 & 51 & 39 & 8.1 &  66 \cr
  \noalign{\medskip\hrule}}} 
         \end{flushleft} 
   \end{table}

We have explored the extent to which subgiant and thick disk star contamination can 
affect the AVR. Because these stars are expected to come from a hotter Galactic 
component, their presence would surely inflate the velocity dispersions of the oldest bins, 
making the AVR artificially steeper. Although our adopted metallicity cut [Fe/H] $< -0.5$ 
already prevents our AVRs from being severely contaminated by thick disk stars, a proper 
separation between the Galactic components can only be made with both kinematical and chemical 
data (Nemec \& Nemec \cite{nemec}; Ojha et al. \cite{ojha}; Chiba \& Beers \cite{chiba}). 
This is a classical problem of mixture decomposition that can be reasonably addressed with 
a multivariate finite mixture analysis. We have used the {\tt EMMIX} 
program\footnote{The program {\tt EMMIX} is freely distributed for non-commercial use at  

{\tt http://www.maths.uq.edu.au/\char126 gjm/emmix/emmix.html}.} (McLachlan et al. 
\cite{emmix}), which fits normal mixture models to multivariate data, using maximum likelihood 
through the Expectation-Maximization algorithm (Dempster et al. \cite{dempster}). An initial group 
configuration must be provided to the program, from which it looks for the best partition amongst 
the data. The best partition is not very much sensitive to this initial guessing, which can be 
either a random grouping or a pre-grouping using a K-means clustering (Hardigan \cite{hardigan}). 
The number of groups, $n$, among which the data will be partitioned must also be provided as 
input. Non-parametric tests (e.g., Tarter \& Lock \cite{tarter}) or physical considerations 
can be used to select $n$. Since we are mostly interested in singling out eventual thick-disk 
contaminators, we will assume $n=2$. We applied the partitioning algorithm to our initial sample 
of 424 stars, in the four-dimensional space defined by [Fe/H], $u$, $v$ and $w$. The average 
properties of the two best-partitioned groups are shown in Table~\ref{partition}. 
The statistical decomposition yields a young, kinematically-cold, metal-rich group and a 
predominantly old, kinematically-hot, metal-poor group. It is reasonable to identify this later 
group with a different, contaminant population, although its $\overline{\rm [Fe/H]}$ is somewhat 
lower than typical thick disk metallicities, suggesting that old thin disk stars may have 
interloped this group. {\tt EMMIX} outputs the a posteriori probability, $p_i$, of each star being 
member of  group $i$. To avoid thin-disk interlopers, we have considered a likely contaminator 
every star with $p_{\rm II} > 0.75$. After discarding these stars from the sample, the AVR 
was calculated with and without the $\vert W\vert$ weighting, using the law $\sigma \propto t^x$. 
The resulting exponents can be found in Table~\ref{heatlaw}, together with a summary of all other 
fitted AVRs. We have found that the presence of thick-disk and subgiant contaminators in the 
sample can increase the exponent of the power-law AVR by up to 40\%, in all velocity components. 
This caveat must be taken into account in every AVR investigation, especially 
in face of the growing evidence supporting the singularity of the thick disk 
(Robin et al. \cite{robin}; Gratton et al. \cite{gratton}).

    \begin{table*} 
      \caption[]{Fitted age--velocity dispersion relations.} 
         \label{heatlaw} 
         \begin{flushleft} 
    {\halign{%
    $#$\hfil &  \quad\hfil #\hfil & \quad\hfil #\hfil  &  \quad\hfil #\hfil  & \quad\hfil $#$\hfil &  
     \quad\hfil $#$\hfil   
    & \quad\hfil $#$\hfil & \quad\hfil $#$\hfil &  \quad\hfil # \hfil \cr  
    \noalign{\hrule\medskip} 
   {\rm parametrization} & binned & age & $\vert W\vert$ & x_{\rm tot} & x_u & x_v & x_w  & note \cr
                         & by & calibration & weight &  &  &  &  & \cr
    \noalign{\medskip\hrule\medskip} 
\sigma\propto (1+t/\tau)^x & number & RPM98 & no & 0.27\pm 0.01 & 0.24\pm 0.02 & 0.30\pm 0.03 & 0.34\pm 0.02 & \cr
\sigma\propto (1+t/\tau)^x & number & RPM98 & yes & 0.50\pm 0.02& 0.47\pm 0.02 & 0.57\pm 0.04 & 0.46\pm 0.03 & \cr
\sigma\propto t^x         & age    & RPM98 & no & 0.42\pm 0.04 & 0.40\pm 0.02 & 0.48\pm 0.04 & 0.41\pm 0.03 & \cr
\sigma\propto t^x         & number & RPM98 & no & 0.26\pm 0.03 & 0.23\pm 0.04 & 0.28\pm 0.04 & 0.35\pm 0.06 & \cr
\sigma\propto t^x         & number & Donahue (1998) & no & 0.33\pm 0.05 & 0.29\pm 0.05 & 0.39\pm 0.05 & 0.37\pm 0.08 & \cr 
\sigma\propto t^x         & number & Donahue (1998) & yes & 0.50\pm 0.08 & 0.36\pm 0.09 & 0.49\pm 0.13 & 0.53\pm 0.07 & \cr
\sigma = a_0 + a_1 t^x    & number & RPM98 & no & 0.24\pm 0.04 & 0.23\pm 0.04 & 0.37\pm 0.07 & 0.30\pm 0.09 & \cr
\sigma\propto t^x         & age    & RPM98 & yes & 0.24\pm 0.06 & 0.24\pm 0.07 & 0.19\pm 0.05 & 0.25\pm 0.03 & $p_{\rm II} < 0.75$ \cr
\sigma\propto t^x         & age    & RPM98 & no & 0.24\pm 0.04 & 0.25\pm 0.06 & 0.28\pm 0.04 & 0.25\pm 0.03 & $p_{\rm II} < 0.75$ \cr
  \noalign{\medskip\hrule}}} 
         \end{flushleft} 
   \end{table*} 

In Figs. \ref{agevelc}e and f, we present the value of the ratio between the semiaxes of the velocity 
ellipsoid, $\sigma_v:\sigma_u$ and $\sigma_w:\sigma_u$. The values we have found are closer to 
those in Wielen (\cite{wie74}) and in Cayrel de Strobel (\cite{cayrel}), as well as those found by 
Dehnen (\cite{dehnen}), for the 14369 HIPPARCOS stars. The axial ratios seem to be nearly constant, 
in opposition to the findings by Knude et al. (\cite{knude}), according 
to which the young groups have a velocity ellipsoid with a substantially high $\sigma_v:\sigma_u$, 
that decreases for old groups. This behaviour, as well as the very high values for 
$\sigma_w:\sigma_u$ found by Meusinger et al. (\cite{meusinger}) could reflect basically selection 
effects in the data samples. Finally, in  panel {\it g} we show the asymmetric drift, which is the trend 
of the rotational velocity of a stellar population lagging behind more and more slowly 
the LSR, the larger its velocity dispersions (G\'omez \& Mennessier \cite{gomez}; 
Binney \& Merrifield \cite{merrifield}).

In spite of the apparent constancy of the velocity axial ratios, a sudden jump is seen in both 
$\sigma_v:\sigma_u$ and $\sigma_w:\sigma_u$ at 1.2 Gyr ago, amongst our data as well as in the 
analysis of Wielen (\cite{wie74}), Knude et al. (\cite{knude}) and Meusinger et al. (\cite{meusinger}). 
We presently do not have an explanation for this. If this feature is real, as these different analyses 
suggest, this could be an important constraint to the AVR theory.


\section{Vertex deviation}

The stars show a velocity distribution that can be characterized by a velocity ellipsoid 
(Schwarzschild 1907; Mihalas \& Binney 1981), with semiaxes $\sigma_{u1}$, $\sigma_{v1}$ and 
$\sigma_{w1}$, that correspond to the  principal velocity dispersions of the stellar group. The 
semiaxes of the ellipsoid generally do not correspond to the coordinate axes used for the 
calculation of the spatial velocities (i.e., $u$, $v$ and $w$), and some calculations must be 
done for the determination of these axes. In practice, it is known that the semiaxes $\sigma_{u1}$ 
and $\sigma_{v1}$ are in the galactic plane (for instance, see Dehnen \& Binney \cite{dehbin}), and the 
problem is restricted to the determination of the angle $\psi_V$ that indicates the galactic 
longitude towards which the principal major axis of the velocity ellipsoid points. This 
angle is called vertex deviation. Once determined, the peculiar velocity dispersions can be 
calculated with respect to the semiaxes of the velocity ellipsoid.

The vertex deviation $\psi_V$ is defined by 
\begin{equation}
\psi_V = \frac{1}{2}
\arctan{\frac{2\sigma^2_{uv}}{\sigma^2_u-\sigma^2_v}} \ ,
\end{equation}
where $\sigma^2_{uv}$ is the covariance between $u$ and $v$. 
We have used a larger bin (each 
having 80 stars) to increase the significance of the results.

      \begin{figure}
      \resizebox{\hsize}{!}{\includegraphics{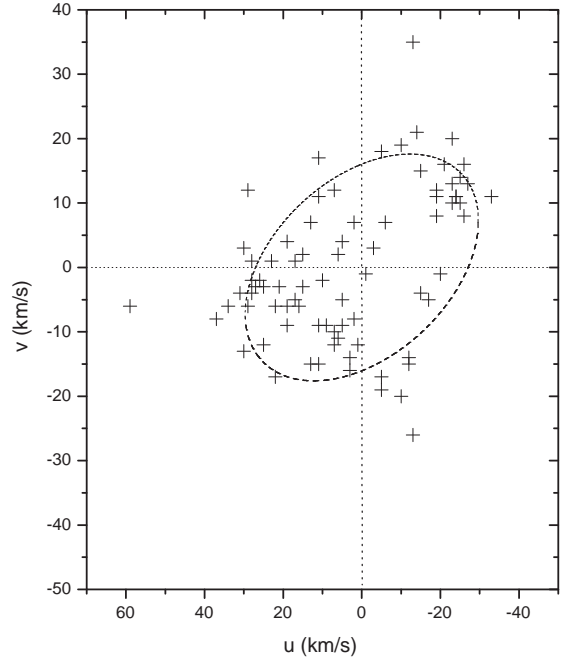} }
      \caption[]{ 
        Projection of the velocity ellipsoid of the stars younger than 1 Gyr onto the $uv$-plane. The 
        tilted ellipsis with respect to the coordinate axes illustrates the vertex deviation 
        of the data points.
          }
      \label{vertex1}
      \end{figure}

Figure \ref{vertex1} presents the data points used for the youngest age bin used in this analysis. The 
dashed ellipsis is the projection of the velocity ellipsoid onto the $uv$-plane. The vertex 
deviation amongst these youngs stars is clearly seen by the tilt of the ellipsis with 
respect to the coordinate axes.

The classical behaviour for the vertex deviation can be found in Delhaye (\cite{delhaye}). 
The vertex deviation is negative ($-$50$\degr$) amongst O and B stars, but positive ($\sim 30\degr$) 
amongst young dwarfs (mainly A type stars). Later dwarfs also present positive vertex 
deviations, but considerably smaller. The vertex deviation for very red and old stars is quoted as 
nearly zero. This relation between vertex deviation and spectral type have led many 
investigations to assume an implicit relation between the age of the stellar group 
and its vertex deviation (e.g., Oblak \cite{oblak1}; Oblak \& Cr\'ez\'e \cite{oblak}), since later dwarfs have 
a higher average age than early dwarfs. 

Surprisingly, we do not find a similar behaviour in our data, as can be seen from Table~\ref{vertex}. 
Either using all the stars in our sample or just those with vertical velocity smaller than 
15 km/s, there is weak indication of an evolutive trend in the vertex deviation, 
if any. Moreover, the greatest vertex deviation is not seen amongst the youngest stars.

Almost all of the former studies in the literature that tried to measure the dependence of the vertex 
deviation on the age used groups of stars with several ages. An average age for the group was 
calculated from the observed properties of stars, and this average age was used in the derivation of 
the function $\psi_V(t)$. The main difference between this approach and ours is that the former can 
only measure integrated kinematical properties of a stellar group (as a function of spectral type or 
colour). They are cumulative kinematical properties. In our approach, since we have ages for 
individual stars, we can consider the kinematical properties in each age bin, in a differential 
approach.

     \begin{table} 
      \caption[]{Vertex deviation.} 
         \label{vertex} 
         \begin{flushleft} 
    {\halign{%
    \hfil #\hfil & \qquad\qquad\qquad\qquad\hfil #\hfil & \qquad\qquad\quad # \cr
    \noalign{\hrule\medskip} 
       $\langle$Age (Gyr)$\rangle$ & $\psi_V$ & \cr 
    \noalign{\medskip\hrule\medskip} 
\noalign{all stars\medskip} 
0.39 $\pm$ 0.20 & 14.8$\degr$ &   \cr 
1.64 $\pm$ 0.52 & 16.0$\degr$ &  \cr 
3.20 $\pm$ 0.45 & 20.0$\degr$ &  \cr 
5.20 $\pm$ 0.81 & 11.3$\degr$ & \cr 
9.16 $\pm$ 1.81 & 10.8$\degr$ & \cr 
    \noalign{\medskip\hrule\medskip} 
\noalign{stars with $\vert W\vert\le$ 15 km/s \medskip} 
0.40 $\pm$ 0.20 & 17.5$\degr$ & \cr
1.58 $\pm$ 0.51 & 9.1$\degr$ & \cr
3.28 $\pm$ 0.44 & 18.6$\degr$ & \cr
5.22 $\pm$ 0.88 & 14.5$\degr$ & \cr
8.47 $\pm$ 1.34 & 0.6$\degr$ & \cr
    \noalign{\medskip\hrule\medskip} 
\noalign{stars with $e \le$ 0.15 and $\vert W\vert\le$ 15 km/s \medskip} 
0.46 $\pm$ 0.18 & 23.4$\degr$ & \cr
1.57 $\pm$ 0.51 & 14.0$\degr$ & \cr
3.20 $\pm$ 0.45 & 4.9$\degr$ & \cr
5.18 $\pm$ 0.76 & $-$13.3$\degr$ & \cr
9.23 $\pm$ 2.06 & 9.5$\degr$ & \cr
    \noalign{\medskip\hrule\medskip} 
\noalign{stars with $\sqrt{u^2+v^2} \le 40$ km/s and $\vert W\vert\le$ 15 km/s \medskip} 
0.39 $\pm$ 0.20 & 16.4$\degr$ & \cr
1.45 $\pm$ 0.49 & 19.6$\degr$ & \cr
3.26 $\pm$ 0.45 & $-$1.5$\degr$ & \cr
5.12 $\pm$ 1.68 & $-$9.2$\degr$ & \cr
  \noalign{\medskip\hrule}}} 
         \end{flushleft} 
   \end{table} 

There are few studies in the literature that do use a differential approach for the study of the vertex 
deviation. Their results give some support to our findings, regardless of their authors conclusion 
on the contrary. Byl (\cite{byl}) has used evolutionary isochrones to calculate 
average stellar ages of nine stellar groups ranging from 0.38 to 10.50 Gyr old. The evolution of 
the vertex deviation amongst these groups does not show the implicit smooth trend expected 
from the classical theory, although much of the scatter could be blamed on the small number of stars 
in each of the nine groups. Wielen (\cite{wie74}) has also found high values for
$\psi_V$ at advanced ages, and a variation uncorrelated with age. Moreover, Wielen has also 
argued that the near zero vertex deviation found amongst late-type dwarfs could be a selection 
effect: Classic vertex deviation studies usually limited the samples to those stars with 
orbital eccentricity smaller than a given value, to avoid the contamination of the sample with 
high-velocity stars. By doing this, they sample the phase space within an ellipsis with 
semi-axes pointing to the main Galactic plane axes, thus biasing the sample. For old, late-type 
stars, which have large velocity dispersions, this bias would favour low $\psi_V$ measurements. 
By using a less biasing selection criteria, $\sqrt{u^2+v^2} \le 40$ km/s and $\vert W \vert 
\le 15$ km/s, Wielen showed that $\psi_V$ in the range 10$\degr$ to 
30$\degr$ could be found amongst stellar groups thought to have nearly zero deviation. We have tested 
both selection criteria with our sample. The vertex deviation amongst our stars using the classical 
selection criteria (orbital eccentricity $e \le$ 0.15 and $\vert W \vert \le 15$ km/s) and after 
applying Wielen's selection criteria are given in Table \ref{vertex}. 
These data were calculated within each of the original bins, using only the stars that 
passed the selection criteria. Since the number of stars passing the Wielen criteria at the oldest age 
bin was very small, we combined the two oldest age bins. 
The result is somewhat close to the expected bias pointed by Wielen. However, it is clear that the 
value $\psi_V$ varies substantially depending on the selection criteria used. Same order variations 
in the measured $\psi_V$ have also been reported when the sample is limited by a coordinate 
range (Oblak \cite{oblak1}; N\'u\~nez \& Figueras \cite{nunez}). 

This variation could be caused by statistical errors in the calculation of $\psi_V$. If we consider 
only the error propagation from the velocity errors (taking as $\approx \pm 2$ km/s for each star), 
through a Monte Carlo simulation, we have $\epsilon_{\psi_V} \approx 0.4 \degr$. But the error also 
depends on the Poisson noise due to sample size. We have run a bootstrap routine to randomly select 
stars in each bin (so that some stars were counted twice or more and some were neglected from a 
given bin) and calculated the vertex deviation. This bootstrap selection was performed 1000 times. 
We have found that the average error in $\psi_V$ is $\la 7.5 \degr$. This indicates that statistical 
errors alone cannot explain the variations we have found.

Even non-differential analyses have not made a strong case for a smooth vertex deviation--age 
relation. The data compiled by Mayor (\cite{may72}) shows a large scatter in the value of $\psi_V$ 
for each, adopted coeval, stellar groups. Particularly interesting is the fact that active HK dwarfs 
in his sample present a smaller vertex deviation (12$\degr$) than inactive HK dwarfs (20$\degr$), a 
result that bears resemblance with the entries of Table \ref{vertex}. Mayor concludes that 
large vertex deviations are observed amongst stars of all ages, provided that their orbital eccentricity 
is low. Recently, Ratnatunga \& Upgren (\cite{ru97}) used a maximum-likelihood analysis in a sample of 
more than 700 Vyssotsky K and M dwarfs and showed that their vertex deviation is significantly non-zero, 
amounting to $12\degr\pm 4\degr$. 

The existence of a smooth decreasing vertex deviation--age relation could provide an important clue 
about the cause of the vertex deviation. Theoretical work on this subject are generally 
split between those theories that propose the vertex deviation is a reflection of the initial 
conditions, which we will call {\sl congenital} scenario, and those, which explain it as the effect 
of perturbations in the local Galactic potential, here called {\sl perturbation} scenario. 

Congenital scenarios proposed include the natural evolution of the original stellar configuration 
at birthplace taken as a spiral arm (Wooley \cite{wooley}; Yuan \cite{yuan}), the launching of 
stars recently formed at the spiral arms with postshock velocity (Hilton \& Bash \cite{hilton}) 
or emerging from an expanding shell (Moreno, Alfaro \& Franco \cite{moreno}), a temporary arrangement 
in the phase space due to the epicycles (Byl \& Ovenden \cite{bylovo}; Bassino, Dessaunet \& Muzzio 
\cite{bassino}) or oscillations in the stellar orbits (House \& Innanen \cite{house}) and 
the superposition of macroscopic motions over the Galactic rotation (Kato \cite{kato}). 
Perturbation scenarios address the effects of non-axisymmetry in the Galactic potential 
(Sanz \& Catal\'a \cite{sanz}; Cubars\'{\i} \cite{cubarsi}) and gravitational perturbations 
caused by the spiral arms (Mayor \cite{may70}, \cite{may72}).

So far, none of these hypotheses can explain satisfactorily the vertex deviation phenomenon 
in all spectral classes. Some of the explanations are particularly suitable for the earlier 
stars, like the launching of recently formed stars by spiral arms or expanding shells, while 
other are more appropriate to explain the deviation found amongst late-type stars.  

A third class of explanation has gained momentum in the last years. The substantial improvement 
in the observational techniques has allowed the investigation of the velocity phase space of the 
solar neighbourhood with unprecedented resolution. These studies have showed very convincingly 
that the velocity phase space is not homogeneous, but punctuated by small scale structures and 
moving groups (e.g., Dehnen \cite{dehnen}; Skuljan, Hearnshaw \& Cottrell \cite{skuljan}). 
Dehnen (\cite{dehnen}) has shown that these moving groups themselves are the main responsible for 
the vertex deviation, and it is easy to understand this: Clumpy structures in the phase space 
puts a large weight on $\sigma^2_{uv}$, yielding a larger $\psi_V$. This idea was first 
raised by Palou\v s (\cite{palous}), who argues that the Hyades and Sirius superclusters 
contaminate local samples, imprinting on their velocity distribution the vertex deviation. 
Moreno et al. (\cite{moreno}) present a similar reasoning for the explanation of the negative 
vertex deviation amongst OB stars.

This could explain why the vertex deviation varies so widely depending on the selection 
criteria used and the coordinate range of the stars used in its determination. These different 
criteria would discard totally or partially some moving groups from the phase space, and this 
could induce to uncorrelated variations of $\sigma^2_{uv}$. In this case, the vertex deviation 
problem would be reduced to explaining the surviving of old moving groups and their peculiar velocities.  
A very original and promising idea was introduced by Dehnen (\cite{dehnen00}) and further developed 
by Fux (\cite{fux}) and M\"uhlbauer \& Dehnen (\cite{muhl}), according 
to which several features of the velocity phase space can be explained by the kinematic interaction 
between the outer disk and the Galactic central bar.

\section{Orbital parameters and stellar birthplaces}

From its spatial velocities and distance, it is possible to calculate the orbit of a star around the
galactic center. The orbit is characterized by some parameters: perigalactic and apogalactic distance, 
$R_p$ and $R_a$, average galactocentric distance, $R_m$, orbital eccentricity, $e$, and maximum height 
above the galactic plane, $Z_{\rm max}$. When a star is born from a molecular cloud, $e$ and 
$Z_{\rm max}$ are nearly zero. With time, due to the orbital diffusion, both parameters will grow, 
and the stellar orbit will become less circular. The orbits were integrated in a galactic potential, 
consisting of an exponential thin disk, a spherical bulge and a dark halo, as described in Flynn et al. 
(\cite{flynn}). The value for the orbital parameters were calculated for 380 of the 459  stars in the
kinematical sample, which correspond to the number of stars that follow the some additional criteria used 
in Table 2 of Paper I, that is, stars should be closer than 80 pc to avoid substantial reddening in the 
photometric indices, brighter than 8.3 mag in $V$ magnitude, less active than $\log R'_{\rm HK} \ge -4.20$ 
and have nominal chromospheric age lower than 15 Gyr.

Let us consider the problem of the determination of the stellar birthplace. Wielen (\cite{wie77}) 
claims that it is impossible to know with accuracy the galactocentric radius of the stellar birthplace 
from its orbit, due to the diffusion of stellar orbits in phase space caused by local fluctuations 
of the Galactic gravitational field. In each epicycle, the star would be scattered in 
random directions, and this would destroy any information about the origin of the star. 

      \begin{figure}
      \resizebox{\hsize}{!}{\includegraphics{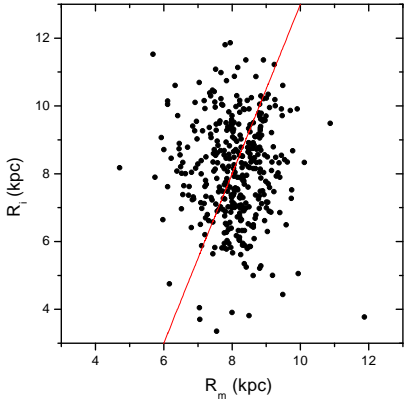} }
        \caption[]{
        Comparison between two proposed birthplace indicators. The solid line shows the prediction 
         by Wielen et al. (\cite{WFD}).
          }
      \label{rmri}
      \end{figure}

Grenon (\cite{grenon}) proposed that, even if the present stellar position cannot be taken as an 
indicative of its birthplace, the average radius of its orbit, $R_m$, is kept close to the initial 
galactocentric radius of the stellar birthplace. This would make possible the use of $R_m$ in the 
derivation of radial constraints, like the abundance gradients and radial variation of the 
age--metallicity relation. 

Wielen, Fuchs \& Dettbarn (\cite{WFD}) criticized this conclusion, pointing that $R_m$ must also be 
considerably modified due to the orbital diffusion. Due to irregularities produced by, for instance, 
giant molecular clouds and spiral arms, the average galactocentric radius of a stellar orbit $R_m$, at 
time $t$, would not be directly linked with the galactocentric radius where the star was born, $R_m(0)$. 
Wielen et al. have described the star scattering process due to the orbital diffusion, and have shown 
that after some Gyr it is impossible to know where the star was born, from $R_m$, with a accuracy better 
than some kpc. Also, to explain the high metallicity dispersion in Edv93 data, these authors suggest 
that the Sun and some other nearby stars were born closer to the galactic center, and have migrated 
to their present location due to the orbital diffusion. Supposing a very well mixed ISM, the authors 
propose that the stellar birthplace, $R_i$, can be roughly estimated from the age and metallicity of 
a star, by the equation
    \begin{equation} 
 R_i-R_\odot \approx -11{\rm [Fe/H]} -0.53t_9 +0.6, 
       \label{ri} 
    \end{equation} 
where $R_\odot$ is the present solar galactocentric radius, and $t_9$ is the age of the star, in Gyr. 
According to Wielen et al., there is a relation between $R_i$ and $R_m$, given by
\begin{equation}
         \langle R_m - R_\odot \rangle = 0.4 \langle R_i - R_\odot \rangle.  
         \label{rmri2}
\end{equation}

More recently, Nordstr\"om et al. (\cite{bruxa}) have argued that the predictions by Wielen et al. 
may be overestimated and that $R_m$ seems to be a good indicator of the stellar birthplace.

      \begin{figure*}
      \resizebox{\hsize}{!}{\includegraphics{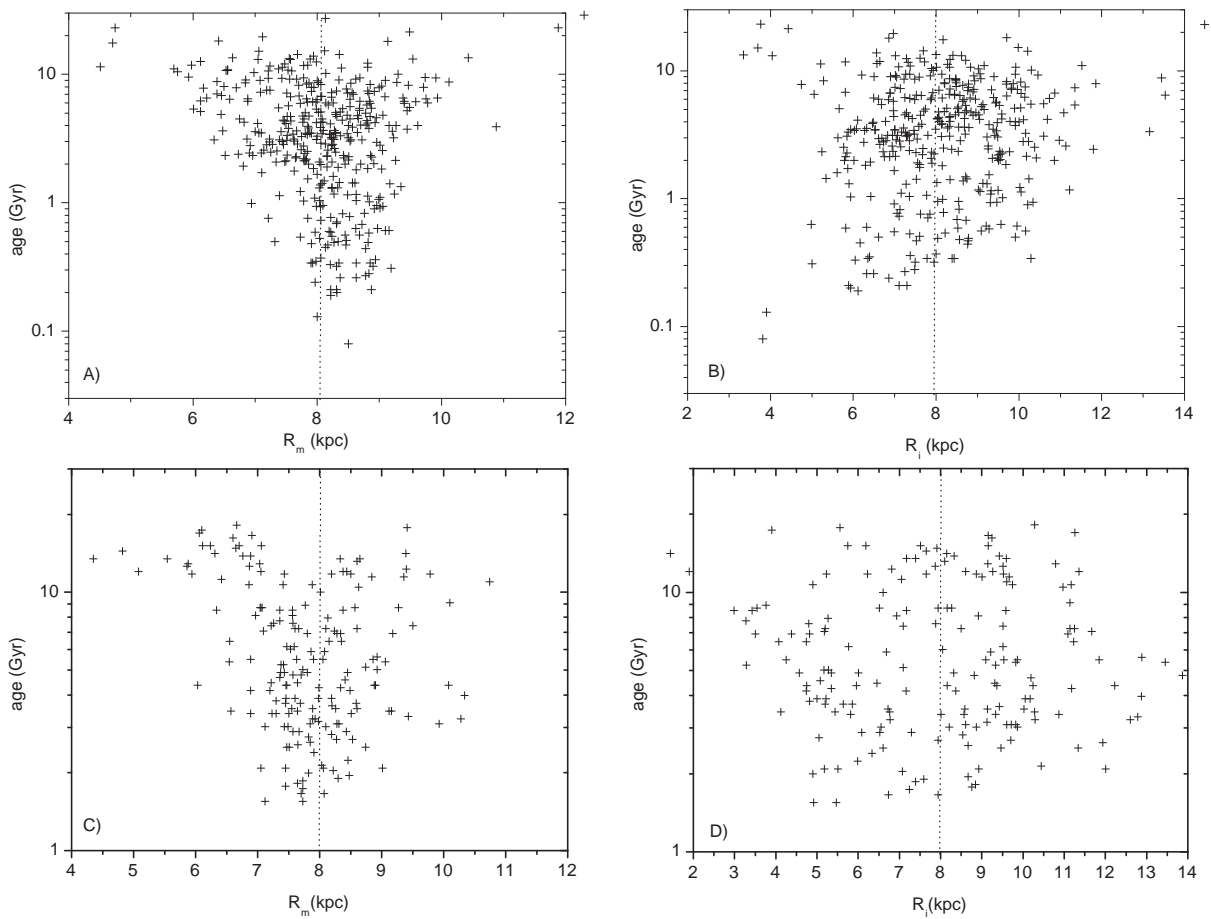} }
       \caption[]{
        Correlation between two proposed birthplace indicators. Panel a, 
       $R_m$ versus age for our sample; panel b, $R_i$ versus age for our sample; panel c, 
       $R_m$ versus age for the sample of Edv93; panel d, $R_i$ versus age for the sample of 
       Edv93. The dashed line indicates the present galactocentric radius of the Sun.
          }
      \label{rmage}
      \end{figure*}

Our sample can be used to test whether $R_i$ and $R_m$ present the same behaviour predicted by 
Wielen et al. (\cite{WFD}). In Fig. \ref{rmri} we show a plot with values for $R_i$ and $R_m$. The 
solid line indicates the predictions by Wielen et al. (\cite{WFD}). The scattering is too large to allow 
a significant conclusion. Moreover, the small range of $R_m$ in our sample makes difficult the 
comparison with the theory. 

Nevertheless, if $R_i$ (or $R_m$) are good indicators of the stellar birthplace, we must find a relation 
between these parameters and the stellar age, since a finite time is necessary for a star 
to be able to move from its birthplace to the present galactocentric radius of the Sun.   
Thus, we expect to find stars born at several galactocentric radii only amongst the oldest stars, since 
the youngest would not have had time to migrate radially  1 or 2 kpc 
from its initial position.

In Fig. \ref{rmage}, we show how these parameters are linked with age. Panel a presents $R_m$ 
as a function of the stellar age, while panel b presents $R_i$. The data are for our sample. 
$R_m$ presents a well-marked correlation with age. It is clear that young objects have all $R_m\approx 
R_\odot$, and that older stars present a higher proportion of objects coming from different galactocentric 
radii. If we assume that $R_i$ represents the stellar birthplace radius, we do not find the same behaviour. 
It is difficult to understand how a star with less than 1 Gyr could come from galactocentric radii much 
smaller than $R_\odot$. Panels c and d present the same comparisons, for the sample by Edv93. The same 
conclusion can be taken from them. The comparison for $R_i$ is even worse than for our sample. 

This figure shows that Eq. (\ref{ri}) is not suitable to calculate the stellar birthplace radius. However,
it does not imply that the outwards migration of the Sun has not occurred. The formalism by Wielen et al. 
is developed for a well-mixed interstellar medium, in which the metallicity dispersion seen amongst the 
stars is produced by the orbital diffusion. However, if there is instead a non-negligible metallicity 
dispersion in the gas, at each galactocentric radius, the straightforward use of the stellar age and 
[Fe/H] in Eq. (\ref{ri}) will only give a rough estimate for the birthplace radius.

We have also tested this idea with a simulation of the stellar orbits diffusion. We have inserted tracer 
stars in an exponential disk under the 3D model for the Galaxy, as described in Flynn et al. (\cite{flynn}). 
New stars are born each time step, simulating a constant SFR. The stars are given small random kicks 
in their velocities at each time step.  Stars near the solar circle get sufficient kicks for being 
heated up to have values of the velocity dispersion ellipsoid similiar to the observed one. Closer 
to the galactic center, the heating is greater, increasing exponentially inward with the same scale 
length as the disk light distribution (circa 4 kpc), with less heating in the outer disk. Inside 2 kpc, 
the heating is cut off, to simulate less molecular clouds there and to simplify the computation. Indeed, 
such stars are not expected to be deflected onto orbits which eventually take them out to the solar 
circle over the disk lifetime.

At the end of the simulation, we find all stars which are on or near to the solar circle. We have 
computed the mean guiding radius of their present orbit, $R_m$ and compared it with the a priori 
known birth radius $R_i$. The relationship found between them is $\langle (R_i-R_\odot)\rangle 
\approx 2 \langle (R_m-R_\odot) \rangle$ as expected by the theoretical considerations performed 
by Wielen et al. (\cite{WFD}).

The problem does indeed come when one tries to recover the relationship from the data. The only way 
we have of recovering $R_i$ from the data is to use [Fe/H]. The disk gradient is only about 0.1 dex/kpc, 
and the scatter in the local metallicities is (intrinsically) of the same order or larger (e.g., Edv93; 
Paper I). Here, by scatter, we mean the width of the [Fe/H] distribution. In our sample, $R_m$ lies in 
the range 6 to 10 kpc, so that the birth radii of the stars lie in the range 7 to 9 kpc. Over such a short 
range in the disk, the mean metallicity of the stars changes by only 0.1 dex in 
either direction, whereas the intrinsic dispersion (over all ages) is 0.12-0.3 dex. 

      \begin{figure*}
      \resizebox{\hsize}{!}{\includegraphics{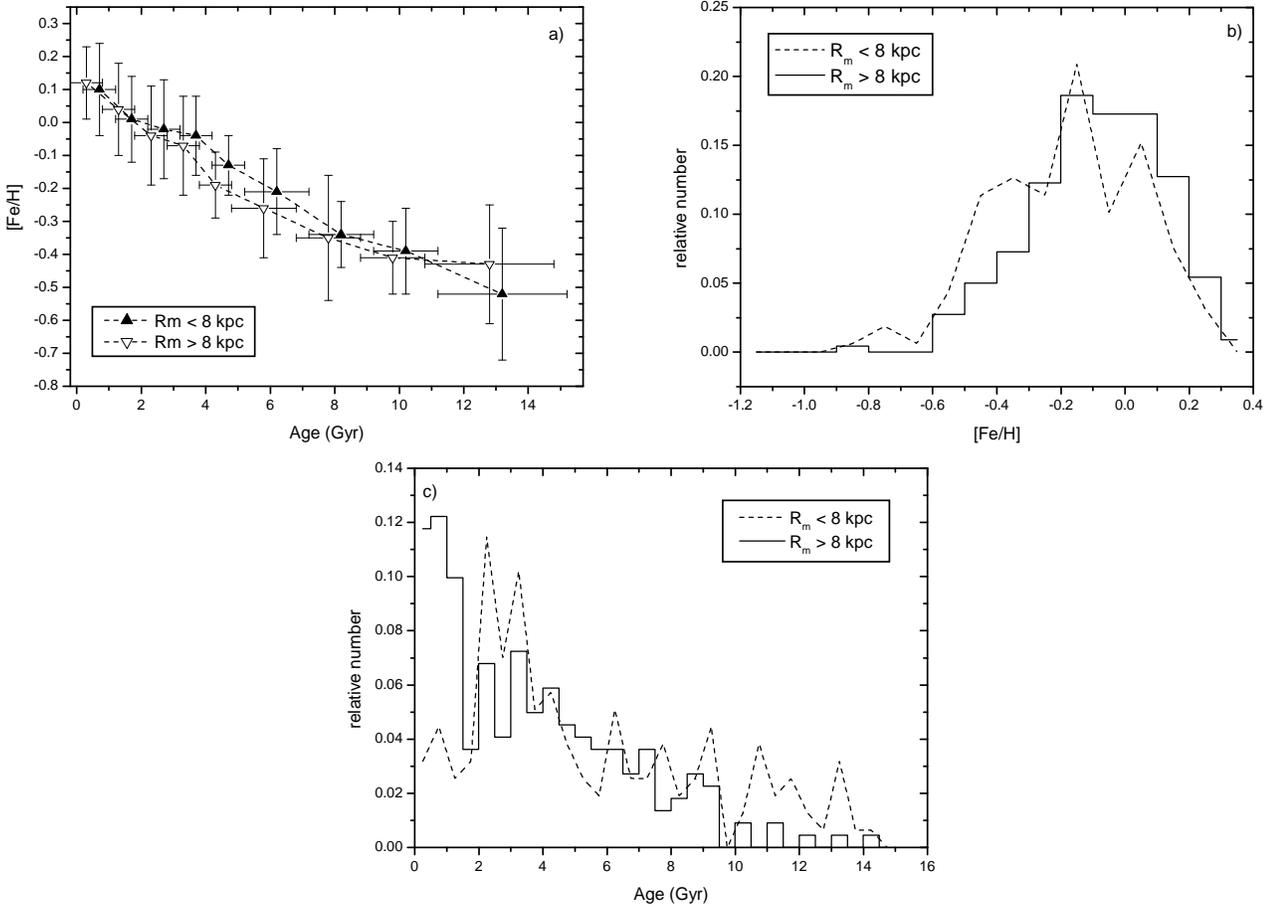} }
      \caption[]{
        Comparison of constraints for two samples, one internal to the solar radius, and the 
        other external to it. Panel 
        a, age--metallicity relation; panel b, metallicity distribution; panel c, age distribution.
          }
      \label{vinkrad}
      \end{figure*}

Our results show that $R_m$ can surely give some information about the birthplace radii 
of the stars. But this information can be used only in a statistical sense, according 
to Eq. (\ref{rmri2}), due to the intrinsic scatter in the $R_m$ relation introduced by the orbital 
diffusion. This must be kept in mind, when interpreting the results from the next section.

\section{Radial constraints to chemical evolution models}

Edv93 have used $R_m$ to investigate the radial variation of some chemical constraints, like the 
abundance ratios and the age--metallicity relation. We decided to use our data sample to derive 
constraints to the age--metallicity relation, the radial variation of the metallicity distribution,  
the star formation rate and the abundance gradient in the disk. 

Due to the narrow range of $R_m$ in our data, we have divided the sample in only two bins according 
to their galactocentric radius:  
stars with $R_m<8$ kpc and stars with $R_m>8$ kpc, which corresponds to stars born internally 
and externally to the solar galactocentric radius, respectively. These groups are composed by 158 
and 222 stars, respectively, adding up to 380 objects. 

Ideally, it should be possible to derive such constraints from a large data sample as ours. However, 
a close look at Fig. \ref{rmage}a shows that there is an intrinsic bias in the $R_m$ distribution. 
The younger objects were born in the vicinity of $R_\odot$, and are badly represented for 
$R_m \neq R_\odot$. Any attempt to derive  radial constraints are affected by this bias.

Let us consider three possible constraints: the age--metallicity relation, the radial variation of 
the metallicity distribution, and the star formation history (from the age distribution), shown in 
Fig. \ref{vinkrad}. At first glance, there is no radial variation in the age--metallicity relation 
(Fig. \ref{vinkrad}a), in opposition to the conclusions by Edv93, that the most inner parts of the 
Galaxy have evolved more rapidly. However, from Fig. \ref{vinkrad}c we see that the sample internal to 
the solar radius does not comprise many young stars, since that have had not time to migrate 
to the solar vicinity. Since the younger stars are generally richer, the radial variation presented in 
this figure cannot be considered real. This explains why the metallicity distribution of the inner 
sample presents a paucity of metal rich stars, compared to the metallicity distribution of the outer 
sample (see Fig. \ref{vinkrad}b}).  The different age distributions in Fig. \ref{vinkrad}c also 
{\it must not be interpreted as radial variation of the star formation history}, because our sample 
does not have an homogeneous $R_m$ distribution as a function of the stellar age.

\begin{figure}
      \resizebox{\hsize}{!}{\includegraphics{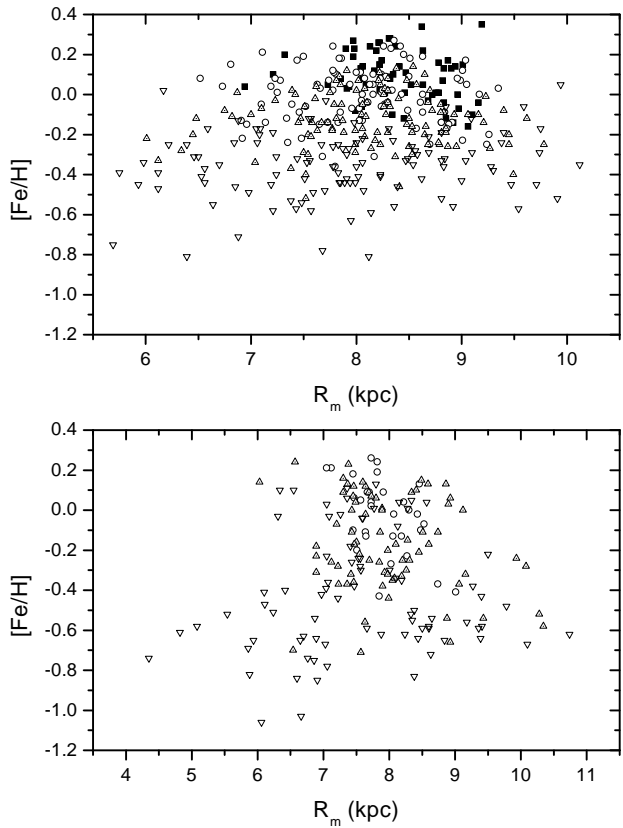} }
       \caption[]{
        Metallicity--$R_m$ diagram. Upper panel: our data; lower panel, 
        Edv93's data. The symbols indicate: filled squares, stars having 
        ages lower than 1 Gyr; open circles, stars aging 1-3 Gyr; dotted 
        triangles, stars having ages from 3 to 6 Gyr; upside down triangles,
        stars with ages greater than 6 Gyr.}
      \label{metrm}
      \end{figure}

In Fig. \ref{metrm} we show the metallicity as a function of the birthplace (upper panel). The different 
symbols separate the objects according to their age, namely: stars with ages lower than
1 Gyr (filled squares), stars with ages of 1--3 Gyr (open circles), stars with ages from 3 to
6 Gyr (dotted triangles), and stars with ages greater than 6 Gyr (upside down triangles). 
We can see that the metal-rich objects are those that were born in the vicinity of the Sun, whereas 
the metal-poor objects come from several galactocentric radii. The same behaviour is present in 
Edv93 data (lower panel). 

The present sample is not particularly useful in the study of radial abundance gradients, in view
of the uncertainties in the determination of $R_m$ and the magnitude of the gradients, which is 
comparable to the average dispersion in  the abundances as taken in the whole range  of galactocentric
radii. However, it is interesting to note that an average [Fe/H] gradient of --0.07 to --0.09
dex/kpc as adopted by Maciel (\cite{m2002}) for open cluster stars fits rather nicely our data
points with ages under 6 Gyr as shown by the filled squares, open circles and dotted triangles
in the upper panel of Fig. \ref{metrm}.

The conclusion by Edv93, that the inner parts of the Galaxy have experienced a more rapid enrichment, 
seems somewhat premature. There is no unbiased coverage in $R_m$ and [Fe/H] in their sample to allow 
such conclusion, just like in our sample. Although radial constraints are extremely needed for the 
understanding of the chemical evolution of the Galaxy, presently there is no sample of late-type dwarfs 
that can be used in their derivation.

\section{Implications for Theoretical Models}

Our results for the AVR, while fairly consistent with independent work, add to the growing evidence
that heating by scattering by GMCs, spiral structures, and accretion, are insufficient to explain
the observations.  Here we summarize the problem, and suggest that a different interpretation
of the AVR should be re-investigated. In the following, we mostly consider velocity dispersions
with respect to a cylindrical coordinate system, where $r$, $\theta$ and $z$ represent the radial,
azimuthal and vertical components of the space velocity, respectively. For Solar neighbourhood stars,
we can adopt $\sigma_w/\sigma_u = \sigma_z/\sigma_r$.

Our major results are:
\begin{enumerate}
\item Excellent power law behavior of all the velocity dispersion components with age, with an 
exponent of about $x$ = 0.24-0.35, similar but somewhat smaller than Wielen's (\cite{wie77}) 
original estimate and some more recent results (Jahrei\ss\ \& Wielen \cite{jahreisswielen}, Fuchs et al.
\cite{fuchsetal}) based on HIPPARCOS data.  Our exponents are increased by about 0.1 if $|W|$
weighting is used, but there is much more scatter in the fits for each component as well as other 
problems described above.  We note that Jahrei\ss\ \& Wielen (\cite{jahreisswielen}) and Fuchs et al. 
(\cite{fuchsetal}) used $|W|$ weighting.
\item  We also find that $\sigma_w$ reaches values around 30 km/s for the oldest stars, while the 
total velocity dispersion reaches 70 km/s.
\item Another constraint on theories for disk heating is the rate of axial ratios of the velocity 
ellipsoid, for which we find $\sigma_w :\sigma_u \sim$ 0.45 to 0.55 in most age bins, with a spread 
of about 0.1 between the different age groups, and little dependence on age except for the two 
smallest age groups. These values are a little smaller than previously adopted values of 0.5-0.6 (e.g.
Wielen \cite{wie77}), but the lack of age dependence agrees with previous work (Dehnen \& Binney 
\cite{dehbin}).
\end{enumerate}

These results appear inconsistent with calculations based on scattering from GMCs and spiral waves, 
primarily because none of these calculations can give a large enough velocity dispersion for old stars.  
The other constraints are more difficult to assess. Lacey (\cite{lacey}) extended Spitzer \& 
Schwarzschild's (\cite{ss1951}, \cite{ss1953}) Fokker-Planck formulation of the problem to 3-dimensional 
disks and found $x \approx 0.25$, $\sigma_z/\sigma_r \approx 0.8$, suggesting that the velocity ellipsoid
was too round. This led to a revival of the idea that transient spiral arms could heat the disk. 
Through resonances of stellar orbits with wave frequencies, primarily radial heating is produced, 
with some of it channeled into the vertical direction. Sellwood \& Carlberg (\cite{sellcar}),
Carlberg \& Sellwood (\cite{carsell}), and Carlberg (\cite{carlberg87}) all found that transient
spiral arms can give $\sigma_z/\sigma_r \approx 0.5$, but obtain inefficient vertical heating,
with $x\approx 0.2$.  This led Jenkins \& Binney (\cite{jb}) and Jenkins (\cite{jenkins}) to analyze 
the problem by combining scattering from clouds and spiral waves.  They found that clouds are indeed 
more efficient vertical heaters in the presence of spiral waves, but still obtained $x < 0.3$.  
Jenkins (\cite{jenkins}) proposed that the addition of disk accretion to the scattering mechanisms 
could alleviate the problem, giving $x\approx 0.4$, but obtained $\sigma_z/\sigma_r \approx 0.65$, 
which appears too round (especially compared to our results). Furthermore, it is not possible to 
reach vertical velocity dispersion of $\sim$ 30 km/s if young stars have $\sigma_w =$ 2-5 km/s 
(Asiain et al. \cite{asiain}; see Kroupa \cite{kroupa}, Fig. 7). Kroupa (\cite{kroupa}) investigated 
the addition of a component due to unbound star clusters which expand at virial or super-virial 
velocities following fast gas removal.  Such a component adds interesting effects to the tail 
of the stellar velocity distribution, but the $\sigma_w$--age relation requires a star formation 
rate that has been strongly declining with time or a cluster mass function which has been shifting 
to less massive clusters with time to the present epoch.

Perhaps the discrepancy is due to the many approximations made in the Fokker-Planck theory, which 
are well-summarized by Villumsen (\cite{villumsen}). Villumsen presented numerical simulations 
that overcame many of these shortcomings.  After a brief initial period of fairly rapid increase 
of heating, the velocity dispersions flatten out to give $x_u \approx x_v = 0.25$ (in agreement 
with Lacey's analytical treatment), and $x_w \approx 0.31$. By the end of the simulations 
($\sim 5\times10^9$ yr) the ratios of velocity dispersions were $\sigma_r : \sigma_\theta : 
\sigma_z = 1.0:0.7:0.6$. Thus the velocity dispersion is too round, although not as round as 
found by Lacey's (\cite{lacey}) analytic result.  Judging from the papers cited above, transient 
spiral arms would reduce  the $\sigma_z/\sigma_r$ ratio, but the addition of disk
accretion should increase it again, by an unknown amount.  Extrapolating Villumsen's
velocity dispersions to $10^{10}$ yr indicates that the total dispersion will not exceed 40
km/s, while $\sigma_z$ will only reach $\sim$ 15 km/s, compared to about 70 and 30 km/s, 
respectively, found here for the oldest stellar group. A major approximation made by Villumsen 
(\cite{villumsen}) is that the number of clouds in the simulations was greatly reduced to 200 
for computational reasons while the masses of the clouds were reduced by the square root of 
this factor to compensate.  It is not known how the diffusion argument on which this scaling 
is based might affect the results.  However, recent high-accuracy simulations of 
$10^6 M_\odot$ GMC heating by some of us (H\"anninen \& Flynn \cite{hahn}) give an even weaker 
increase of the velocity dispersion with time, although the velocity ellipsoid is flatter.

The situation may actually be worse than this.  In a pair of little-cited papers, Yasutomi
\& Fujimoto (\cite{yafu89}, \cite{yafu91}) presented a significantly more sophisticated
treatment of the star-cloud scattering problem, including the non-Keplerian potential of 
individual clouds (assumed Plummer), the nonlinear part of the background disk gravitational 
field (neglected in earlier work which assumed harmonic oscillation through the linear 
epicyclic approximation), the galactic shear of stars and gas clouds, and the non-uniform 
distribution of massive molecular clouds in the Galactic plane (annulus-like 4-8 kpc ring and
spiral-like distributions), but restricted to two dimensions.  The main approximation made is
that stars only interact with their nearest gas cloud. The major effect discovered
was that statistically anisotropic encounters (direct and catching-up encounters)
lead to important systematic accelerations and decelerations of the stars, corresponding to 
dynamical friction.  This leads to very slow growth of the velocity dispersion for times 
$\ge$ 1 Gyr, with $x\approx$ 0.20 for a uniform distribution of clouds, and even 
smaller values $\sim$ 0.10 to 0.15 for clouds distributed in an annulus or spiral arms.  
In all cases the maximum total velocity dispersion was only 30 to 50 km/s, a value reached 
quickly ($\le$ 1 Gyr), before dynamical friction effects become important.  These results 
led Yasutomi \& Fujimoto to suggest that if a scattering process is to account for the
observations, either gas clouds must have been more massive and numerous billions of years
ago (Semenzato \cite{semezato}), and/or there must exist a high surface density of
massive dark objects, such as the $10^6 M_\odot$ black holes proposed by Lacey \&
Ostriker (\cite{laceyostriker}).  The generalization of these calculations to three 
dimensions by Yasutomi \& Fujimoto (\cite{yafu91}) showed that the general results are robust.  
In none of their models did the total velocity dispersion reach values in excess of 35 km/s, 
and in fact slowly {\sl decreased} with time due to the dynamical friction effect, after 
the initial transient heating episode for models with a uniform spatial distribution of clouds.  
Furthermore, they quote a result for the velocity ellipsoid $\sigma_r : \sigma_\theta : \sigma_z
= 1.0:0.7:0.7$, clearly too round compared to the results found here or in earlier work.

Kokubo \& Ida (\cite{kokubo}) reexamined the problem using direct integrations
of epicycle orbits to obtain probability distributions for the change in orbital
parameters, but obtained results similar to the other methods. After an initial
rapid increase, the later phases gave $x = 0.25$ for all three velocity
dispersion components (smaller than our result) and $\sigma_z/\sigma_r \approx 0.6$
(slightly larger than our result).  For GMC masses of $10^6 M_\odot$, the
total velocity dispersion only reaches about 30 km/s while $\sigma_w$ only
reaches 10 km/s after 10$^{10}$ yr.  Kokubo \& Ida find that if the GMC
mass is increased to $10^7 M_\odot$, $\sigma_{\rm tot}$ increases to about 50
km/s, while $\sigma_z$  reaches about 20 km/s.  Kokubo \& Ida's claim to the
contrary, these values are still significantly smaller than found here and elsewhere.

Thus the current situation is that {\sl none} of the above investigations of
scattering/heating models can account for either $x$, $\sigma_z/\sigma_r$
or the total or vertical velocity dispersions found here and elsewhere for
older stars, even when shear motions are included.  Including more physical
realism has apparently only exacerbated or accentuated the problems in earlier work.

The ratio $\sigma_z/\sigma_r$ has also been estimated for at least two other disk galaxies,
NGC 488 (Gerssen, Kuijken \& Merrifield \cite{gkm98}) and NGC 2985 (Gerssen, Kuijken \&
Merrifield \cite{gkm00}).  The values that they found ($0.70\pm 0.19$ and $0.85\pm 0.1$,
respectively) are larger than found in the solar annulus, although still significantly less
than unity.  The existence of a broad range in this ratio among disk galaxies would significantly
complicate the interpretation, especially since there are several different mechanisms for producing
a ratio of a given value and uncertainty in the result for each mechanism.  We therefore concentrate
on the absolute values of the velocity dispersions themselves, and confine our interpretation to
the solar annulus where the data is probably most accurate.

One possibility is that the lifetimes of the clouds were assumed to be too large (infinite).  
Fujimoto (\cite{fujimoto}), using a significantly different stochastic approach to the problem 
(theory of random walks in the field of a harmonic oscillator), assumed that GMC lifetimes are small,
$\sim10^7$ yr.  At the time of this work, the large lifetimes ($\sim 3 \times 10^8$ yr) suggested 
by Solomon, Sanders \& Scoville (\cite{solomon}), based on coagulation requirements, were popular, 
and so several papers dismiss Fujimoto's calculations as unrealistic because of the small adopted
cloud lifetimes.  However subsequent work has strongly suggested smaller lifetimes about equal 
to what Fujimoto adopted for ``GMC"s (see Elmegreen \cite{elmegreen}; Hartmann, Ballesteros-Paredes 
\& Bergin (\cite{ballesteros}), and references therein).  The effect of the short lifetimes is 
significant because it provides a strong fluctuating force component which is missing
in all other work.  Kroupa (\cite{kroupa}), unaware of Fujimoto's work, independently  pointed 
out how the small cloud lifetimes would add a fluctuating acceleration to stars because they 
experience a time-varying potential on the timescale of their passage by each cloud.  However,
Fujimoto obtained total velocity dispersions and vertical velocity dispersions of only 40 km/s 
and 6 km/s after $10^{10}$ yr.  This result must be partly related to Fujimoto's small assumed 
GMC mass of $2 \times 10^5 M_\odot$, but the scaling of the heating rate with GMC mass derived
by Kokubo \& Ida (\cite{kokubo}) suggests that even this short-lifetime effect will be insufficient 
to provide the large velocity dispersions of the older stars.

In order to use heating to account for the AVR, it may be necessary to invoke very massive 
black holes (Lacey \& Ostriker \cite{laceyostriker}).  Simulations by H\"anninen
\& Flynn (\cite{hahn}) show that if the total mass of the halo consists of $10^7 M_\odot$ 
black holes, the disk could be heated from a present velocity dispersion of 18 km/s to 
80 km/s in about 8 Gyr (i.e., for an 8 Gyr old population), but if the present velocity
dispersion is 10 km/s, the disk would only be heated to about 60 km/s even for 15 Gyr old 
populations.  Only a combination of GMCs and $10^7 M_\odot$ black holes comprising half 
the halo could heat the disk sufficiently in 12 Gyr, but in that case $\sigma_z/\sigma_r
\approx 0.63$, somewhat larger than found here, although marginally consistent with other results.

Another mechanism which could plausibly result in a velocity dispersion-age relation is the 
stochastic heating due to minor mergers of the Galactic disk with satellite galaxies or passage 
of dark matter substructure through the disk. The dynamical friction force exerted by the disk 
converts a fraction of the satellite kinetic energy into vertical disk energy. Older
stars should have suffered more events, either from more satellites and dark matter substructure, 
or repeated orbital passages of an individual object until it is tidally stripped, resulting in 
a velocity dispersion-age correlation. This process has been studied analytically and numerically 
by a large number of authors (see Vel\'azquez \& White \cite{velazquez}, Font et al.
\cite{font}, Taylor \& Babul \cite{taylor}, Benson et al. \cite{benson} and references therein).
Unfortunately most of this work was aimed at the problems of disk disruption, and then matching 
the scale heights of present-day disk galaxies.  The problem is made more difficult by the 
uncertain partitioning of energy deposited in the disk between disk heating and global modes 
such as warping and bars (which themselves could heat the disk).  Font et al. (\cite{font}) used
simulations of disks heated by the ensemble of subhalos in CDM model halos to infer that the disk 
heating rate (as measured by an estimated diffusion coefficient) is less than the rate required 
to match observations of the solar neighborhood, although the discrepancy is not very large, and 
depends somewhat on the adopted distribution of satellite orbits and especially the mass
distribution of the CDM subhalos.  Benson et al. (\cite{benson}) used similar ensembles of halos 
with a wide range of masses to show that the predicted scale height distribution of present-day 
models (i.e. after billions of years of heating) compares favorably with the distribution of 
scale heights determined by Bizyaev \& Mitronova (\cite{bizyaev}) for a complete sample of 60 
edge-on galaxies, but only when some molecular cloud scattering is included to avoid
a low-scale height tail of the distribution.  However they did not give information concerning 
how the velocity dispersion increases with time, its absolute value, or the ratio of vertical 
to radial velocity dispersion. After 4 Gyr of the orbit of a single satellite, the disk vertical 
energy increases by about 10\% to 80\% for various models of Benson et al. (\cite{benson}).  
As a velocity dispersion increase (i.e. square root of energy increase), this is very small 
compared to the factor of 2 or so increase that the Milky Way exhibits over a similar time,
but the number of discrete satellite bodies could be large, at least initially (about 100 
in the Font et al. models).

One problem we see with this model is that tidal stripping by the disk continually decreases 
the flux of bombarding subhalo structures, so one expects that eventually the velocity 
dispersion-age relation should flatten out at ages less than some value, or at least show 
a change of slope due to the decreasing effectiveness of this process at more
recent times.  Such a flattening is seen in the velocity dispersion-age relations shown 
in Font et al. (\cite{font}).  No such feature is seen in the observations presented here, 
supporting their conclusion that heating by merger with subhalos is unlikely to account 
for the relations.

We cannot rule out the possibility that minor mergers could have imprinted sudden changes 
in the slope of the AVR, especially given the cumulative observational evidence that the 
Milky Way faced and is presently facing mergers with some dwarf satellites (e.g., Ibata, 
Gilmore \& Irwin \cite{ibata}; Gilmore, Wyse \& Norris \cite{wyse}; Newberg et al. 
\cite{newberg}; Yanny et al. \cite{yanny}; Rocha-Pinto et al. \cite{RPetal03}; Majewski et al.
\cite{majewski}). It is tempting to ask whether slope changes in the AVR could
be used to recover the past merging history of the Galaxy. Mergers could also trigger 
star formation bursts in the disk, as Paper II suggests. Nevertheless, we have found no 
clear correlation between the star formation history (derived in Paper II) and the AVR.

Similarly, it appears that vertical heating of a galactic disk by bending instability 
(Sotnikova \& Rodionov \cite{sotnikova}), while possibly important due to the relatively 
long saturation timescale of the instability, appears to give significant ``shelves" in 
the velocity dispersion-time relation. These features appear at the time of maximum amplitude 
of the nonaxisymmetric and then symmetric bending, at relatively small times (0.5 to 1.5 Gyr 
in the models of Sotnikova \& Rodionov), followed by a flat velocity dispersion-time 
relation.  Again, such saturation is not seen in the observations, although it is 
uncertain if the mechanism, in conjunction with the other mechanisms discussed, could 
avoid the saturation. However the radial velocity dispersion component actually decreases 
in time due to this mechanism, again suggesting that it may be important, but would need to be
combined with other mechanisms to explain the observations. The results for heating by 
the bending of a bar presented by Sotnikova \& Rodionov are similar (except that the 
radial dispersion increases somewhat with time), with a factor of 2-3 increase in the 
vertical dispersion produced in about 1.5 Gyr.

A very different, but entirely plausible, way for dissipative gas systems to account for 
the AVR that was first proposed by Tinsley \& Larson (\cite{tinsley}).  They showed, in 
the context of Larson's (\cite{larson1976}) two-fluid models for disk galaxy formation 
and evolution, that the gradual dissipation and settling of the gas into a thinner and 
cooler layer naturally results in a reduction of the vertical velocity dispersion from 
about 50 km/s at age 3 Gyr when the disk becomes well-defined, to about 15 to 12 km/s
for ages between 10 and 15 Gyr.  The stellar velocity dispersion just reflects the 
evolution of the combined internal velocity dispersion and (for the older stars) the 
vertical motion of the gas disk.

Advantages of the Tinsley \& Larson dissipation model are:
\begin{enumerate}
     \item The sign of the $\sigma(t)$ correlation and perhaps the
     velocity dispersion for old stars are accounted for fairly naturally,
     without any fine-tuning of parameters (for later work, see below) or
     introduction of, for example, extremely massive scattering entities;
     \item The correlation between velocity dispersion and average metallicity,
     both of which should depend on age, can be accounted for essentially using
     only one physical process, since turbulent dissipation and the star formation
     rate are closely coupled in disk evolution models.  Disk heating mechanisms
     that involve stellar scattering are decoupled from chemical evolution and
     the star formation rate.
\end{enumerate}

The problem with the model is that it is not clear what the predicted functional form of 
the AVR and the corresponding components should be.  In addition, the evolution of the 
gas velocity dispersion is controlled in part by the adopted parameterization of the
star formation rate in terms of density or other variables, which is
problematic because of the notorious uncertainty in such parameterizations.

Later fluid models for the settling of initial hot protogalactic gas into the equatorial 
plane have become more detailed in input physics making use of the chemodynamical 
prescription of galaxy evolution, including different forms of star-formation feedback, 
a number of heating and cooling processes, and an accounting of the multi-phase character 
of the interstellar medium. Although more parameters are naturally required to accommodate the
increased number of processes, their choice is not arbitrary but formulated from plasmaphysical 
and astrophysical studies. It can also be demonstrated that even if the parametrization 
of the star-formation rate is taken arbitrarily in the chemodynamical prescription it will 
install a dependence on density and temperature of the warm gas phase in a very narrow 
range because of self-regulation processes (K\"oppen et al. \cite{koeppen}, \cite{koe98}).
Burkert, Truran \& Hensler (\cite{bth}) provide a useful example, although only the 
vertical dependence is included in their one-dimensional fluid model.  The results are 
very similar to Tinsley \& Larson, with the velocity dispersion decreasing from $\sim$
40 km/s to $\sim$ 20 km/s after about 5 Gyr.  Although they ascribe this decrease to a
decline in the star formation rate (hereafter, SFR) by a factor of 10 due to gas depletion 
(they assumed the SFR varies with the square of the gas density), this is equivalent to  
Tinsley \& Larson's gaseous dissipation effect, since it is the SFR that drives the 
``cloud" or ``turbulent" motions of the gas from which the stars form in both cases.  
Besides the advantages mentioned above, the work of Burkert et al. (\cite{bth}) shows 
that their models can account for the observed vertical distribution of stellar 
density and velocity dispersion.

An extension of this chemodynamical model to two dimensions and more physical processes 
has been presented by Samland, Hensler \& Theis (\cite{sht}). A notable result of the 
Samland et al. models (their Fig.3) is that, although the total (halo + bulge + disk) 
SFR has an early peak at $\sim$ 2 Gyr and then declines by a factor of five or so, 
the disk SFR is actually fairly flat.  As explained by Samland et al., this is 
because the disk has only become a relatively discrete entity after about 7-8 Gyr, while 
the total duration of the model is 15 Gyr.  This slow disk formation in their models can be
traced back to the large cloud collision (i.e. dissipation) timescales adopted.
Even beginning at 7 Gyr, the disk SFR appears to only decrease by a factor
of three or so, suggesting that the velocity dispersion will not decrease
much with time.

      \begin{figure}
      \resizebox{\hsize}{!}{\includegraphics{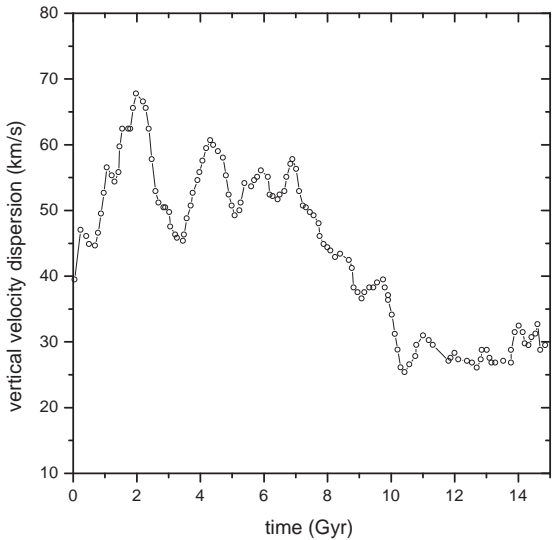} }
        \caption[]{Temporal evolution of the vertical velocity dispersion
     of the cloudy gas phase in the solar vicinity (at 8.5 kpc galactocentric
     distance) as the result of the chemodynamical Milky Way model by Samland
     et al. (\cite{sht}), adapted from Samland (\cite{samland}).
          }
      \label{hensler}
      \end{figure}

Fig. \ref{hensler} shows the time history of the vertical velocity dispersion of the cloudy 
gas phase in the solar vicinity (at 8.5 kpc galactocentric distance) obtained in the chemo-dynamical 
Milky Way model by Samland et al. (\cite{sht}). Since stars are formed from the cool phase this also
represents the velocity dispersion of the newly formed stars.  One can discern that the
vertical dispersion has early oscillations between 45 and 65 km/s, when gas starts to 
accumulate in the equatorial plane and partly cools dissipatively but is still heated 
due to the continuing collapse. Then, after the disk has formed about 8 Gyr ago (in the model), 
the vertical velocity dispersion declines, flattening out at about 25-30 km/s over the 
past 5 Gyr.  For the comparison of this value with observed solar vicinity stars, one has to 
realize that the result is derived as an average over a few vertical grid cells and thus only
represents a thick disk component of almost 700 pc height because of its coarse spatial
resolution.  The ratio $\sigma_z/\sigma_r$ is about 0.5 at most radii in the late-time 
model, in good agreement with our empirical result.

Other models for disk galaxies that include dark matter and cosmologically-motivated
initial conditions have tended to concentrate on chemical evolution and/or high-redshift
properties, and have used SPH + N-body methods rather than fluid models, making it 
difficult to estimate velocity dispersions at late times when small SFRs mean that 
the number of newly-formed star particles is small. However, the
velocity dispersion behavior of such models, when available, are very similar
to those discussed above.  For example, Steinmetz \& M\"uller (\cite{stein95})
state that for stars in age intervals [3,8.5], [1,3], and (indirectly)
$<$ 1 Gyr, the velocity dispersions fall from 80 to 40 to 20 km/s, respectively, at
$R\sim 10$ kpc.  Their Fig. 8b suggests that at late times $\sigma_r > \sigma_\theta \approx
\sigma_z$ over most of the model galaxy that is relevant to the solar neighborhood,
but small number fluctuations and artificial heating by scattering from the
model dark matter particles make the results difficult to interpret.  Once
again, the agreement with the density distribution and chemical properties
(e.g. metallicity gradients, see Steinmetz \& M\"uller \cite{stein94}) of
real galaxies makes this cooling interpretation of the $\sigma(t)$ correlation
attractive from the point of view of ``Occam's razor."

More recent SPH/N-body models of disk galaxies by Berczik (\cite{berczik99},
\cite{berczik00}), which include more stellar components and different
parameterization of stellar energy feedback, as well as a different star formation rate
prescription (basically a virial theorem condition on the SPH particles rather than
either a density dependence or a density threshold) do not give the
time-dependence of the stellar velocity dispersion, but indicates that final values are
$\sigma_z \approx 20$ km/s over the whole disk, with $\sigma_r > \sigma_z$
near the sun (see Berczik \cite{berczik99}, Figs. 6-8).

Thus the time dependence of the disk velocity dependence and its absolute
scale seems robust over a fairly wide range of independent models, agreeing
overall with Larson's (\cite{larson1976}) early work.  It is unfortunate that
the relevant kinematic information from the detailed disk evolution
models of Raiteri, Villata \& Navarro (\cite{raiteri}) and Friedli \& Benz
(\cite{friedli}) are not available to test this robustness further.  The published star
formation rates suggest that the behavior will be similar, although there are larger
apparent temporal fluctuations in the late disk SFRs and presumably in the
stellar velocity dispersions.

However unlike the scattering models, which have great trouble accounting
for the large velocity dispersion of older stars, the cooling models have
a problem in producing small enough velocity dispersions at late times:
Instead of power law cooling, the velocity dispersion approaches a nearly
constant value.  Larson's early models apparently gave the smallest present
day velocity dispersions of 12-15 km/s, while all the other models discussed
above tend to flatten out at about 20-35 km/s.  In fact Tinsley \& Larson
(\cite{tinsley}) and Larson (\cite{larson1979}) were aware of this problem,
and suggested that turbulent dissipation within the disk layer (rather than
settling of the disk itself on larger scales) could lead to the requisite
small velocity dispersion.  The reason this does not occur in the disk
formation models is that at late times the SFR is small, so that the star
formation uses available gas on a long timescale, leading to a relatively
constant SFR.  Since the models tend toward self-regulating equilibrium,
with dissipation balancing SF (see K\"oppen, Theis, \& Hensler
\cite{koeppen}, for one-zone models), the gas and hence stellar velocity
dispersion cannot decrease.

Nevertheless, one reason for the apparent large velocity dispersions in
all the disk formation models at late times might be lack of spatial resolution,
such that the calculation is yielding an average velocity dispersion
over a height much larger than that of the thin disk.

A likely more important aspect of these models is that they all use some
form of sub-grid (unresolved) dissipation whose timescale is essentially
a parameter in the models.  For example, in the chemo-dynamical collapse
models, the dissipation comes through cloud collisions, but the dissipation
timescale is set to about an order of magnitude larger than the cloud collision
timescale based on the geometric cross section.  This allows for slow early
collapse, but also forces the gas to resist a reduction in velocity dispersion
below about 20 km/s.  It is relatively easy to show using one-zone analytic
models that the self-regulated gas velocity dispersion will scale with the
SFR as (SFR)$^{1/3}$ --- see Scalo \& Chappell (\cite{scalochappell}).
In fact a number of independent simulations of interstellar MHD turbulence indicate
that the dissipation time is rapid, of order a crossing time (Mac Low et al.
\cite{maclowetal}; Mac Low \cite{maclow}; Ostriker, Gammie \& Stone
\cite{ostriker}), even in the presence of stellar heating sources
(Avila-Reese \& V\'azquez-Semadeni \cite{avila}).  Thus we ascribe the
inability of collapse models to settle to a small-enough velocity
dispersion to an underestimation of the dissipation rate.  If the dispersion
were really set by an equilibrium between SF and dissipation, this model
would be unable to account for the continued decrease in velocity dispersion
at relatively recent times ($<$ 1 Gyr).  We think the problem is not serious,
since, moving away from one-zone models, more detailed turbulence simulations
show that the turbulence dissipates rapidly in spatial regions between
star-forming events (Avila-Reese \& V\'azquez-Semadeni \cite{avila}),
and new stars will form from this gas.     

In fact it is obvious that this model has no problem in accounting for the low 
velocity dispersion of young stars, since the velocity dispersion of the 
interstellar gas is {\sl observed} to have about the same value as that for 
the stars.  The only problem involves understanding why the existing 
simulations do not dissipate turbulence sufficiently to match the observed 
present-day gas velocity dispersions.

In the future it may be possible to empirically distinguish between the dissipation 
and heating models by comparing the models with observations of $\sigma_z/\sigma_r$ 
in disk galaxies.  Shapiro, Gerssen \& van der Marel (\cite{shapiro}) have used 
major- and minor-axis kinematics to estimate $\sigma_z/\sigma_r$  for a number 
of disk galaxies of various Hubble types. The resulting values of $\sigma_z/\sigma_r$ 
are in the range 0.5 to 0.8 with considerable uncertainties, and are consistent with no
trend with Hubble type at the $1 \sigma$ level, or with a marginally
significant trend of decreasing $\sigma_z/\sigma_r$  with later Hubble type.

Unfortunately, it is not yet possible to test these results with
predictions of the two models.  Simulations by Jenkins \& Binney (\cite{jb})
seem to predict a decrease with increasing molecular gas fraction and
spiral arm strength, so with advancing Hubble type.  The uncertainties
in the empirical ratios are large enough that such a trend cannot be
uncovered at the $1 \sigma$ level.  The galaxy collapse models can produce
$\sigma_z/\sigma_r$ ratios in the observed range, but the inability of the
models to predict the ratio at late times, and the lack of modeling of
disks of different Hubble types, preclude a prediction of any trend.
Hopefully further models of disk kinematics in galaxy formation models
that result in a range of Hubble types will yield such information.
One trend that could in principle be tested is that $\sigma_z/\sigma_r$
varies from 0.6 in the dense inner regions to 0.48 in the outermost
region in the model of Samland et al. (\cite{sht}).  It is not known whether
this variation is generic for the dissipation model.

\section{Conclusions}
     
Following our former investigations on the age-metallicity relation and the star 
formation history, we have used the chromospheric ages for a sample of late-type 
dwarfs to investigate the chemokinematical properties of the solar vicinity. The 
use of chromospheric ages warrants an almost complete coverage of the galactic disk 
history with a sample consisting of stars in the same colour range, something 
that cannot be achieved from other methods.

The exponents for the AVR that we have found frustrantingly seem to span the whole 
range of values which can be used to constrain the theory on the disk heating. 
Nevertheless, after taking into account a statistically more meaningful binning and 
metallicity corrections to the chromospheric age, we arrive at an exponent ranging 
from 0.26 to 0.31, for $\sigma_{\rm tot}\propto (1 +t/\tau)^x$.  

We have shown that the vertex deviation of solar neighbourhood late-type stars 
varied irregularly between the present time and 10 Gyr ago. No clear dependence 
of the vertex deviation on the age of the stellar group was found in our analysis. 
The vertex deviation also varies substantially when different selection 
criteria are applied to the sample. We give support to Palou\v s (\cite{palous}) 
idea that the vertex deviation is caused by the contamination of the sample by superclusters 
and moving groups stars.

Our sample was used to test the idea that there is a statistical relation between 
the average orbital galactocentric distance and the galactocentric radius where 
the star was born. While our simulations provide support to Wielen et al. (\cite{WFD}) 
results, we have found that local data does not have a good coverage in 
orbital galactocentric distance to allow a empirical testing of this relation. Nevertheless, 
the relation between $R_m$ and age that we have found supports Edv93's use of 
$R_m$ as a reasonable indicator of the stellar birthplace radius.

Nevertheless, since young stars born at a galactocentric radius different from $R_\odot$ 
have not had time to scatter into the solar neighbourhood at representative numbers, 
we believe that there is no possibility to find the radial variation of the age--metallicity 
relation or star formation rate using local stars only. A much larger data sample, as that 
expected to be provided by GAIA, could solve this problem. 

A consideration of numerous results of theoretical studies of secular 
heating mechanisms, most involving gravitational scattering from large 
inhomogeneities in the gas or star system, indicates that none of these 
is capable of explaining satisfactorily the large velocity dispersions of the oldest 
disk stars.  The problem is especially serious when one considers the 
dynamical friction effect found by Yasutomi \& Fujimoto (\cite{yafu89}, 
\cite{yafu91}). 

Stochastic heating mechanisms involving satellite interactions are 
possible, but should (and do) give an AVR that flattens out for 
relatively young (few Gyr) stars, a feature that is not observed. 
Instead, we suggest that a very plausible alternative is the one 
suggested long ago by Tinsley \& Larson (\cite{tinsley}), that the velocity 
dispersion of stars just reflects the velocity dispersion of the gas 
from which they formed, and that the turbulent motions of this gas has 
been declining continually since the formation of the disk.  We cite a 
number of more recent disk evolution papers that find similar behaviour. 

While all the models successfully account for the dispersion of the 
older stars, we are unable to say whether the simulations produce the 
low velocity dispersions observed for young stars, due to a combination 
of resolution effects and uncertainties in input physics.  However this 
is obviously only a problem for the simulations, and not for the idea 
itself: the real observed interstellar medium {\sl does} have a velocity 
dispersion that is similar to the velocity dispersion of the youngest 
stars, so we are assured that the model must give the correct result at 
the present time.  Hopefully future studies of disk evolution will 
improve the situation.

There is a straightforward way to test and distinguish the heating and 
cooling models if stellar and gas velocity dispersions for face-on high 
redshift galaxies can be obtained in the future.  The heating model 
predicts that the stellar velocity dispersion should be small at such 
early times, while the cooling models predict that the gas velocity 
dispersion should be large.  We think it unlikely that both conditions 
can occur simultaneously.

\begin{acknowledgements}
      We have made extensive use of the SIMBAD database, operated at CDS, 
      Strasbourg, France. Part of this research was made possible by the use of 
      the multivariate decomposition software {\tt EMMIX}, freely 
      distributed to the scientific community by Geoff McLachlan and collaborators.  
      Markus Samland kindly provided us with data from chemodynamical models. 
      This work was partially supported by FAPESP and CNPq to WJM and HJR-P,
      NASA Grant NAG 5-3107 to JMS, the Finnish Academy to CF and JH, and the 
      Deutsche Forschungsgemeinschaft to GH under grant HE1487/5. HJR-P also 
      acknowledges generous support from Frank Levinson and Wynnette LaBrosse 
      through the Celerity Foundation.
\end{acknowledgements}

\end{document}